\documentclass[aps,reprint,superscriptaddress,nofootinbib]{revtex4-2}
\usepackage[utf8]{inputenc}
\usepackage{amsmath}
\usepackage{amssymb}
\usepackage{bm}
\usepackage{graphicx}
\usepackage{graphics}
\usepackage{color}
\usepackage{textcomp}
\usepackage[bookmarks=false]{hyperref}
\usepackage[usenames,dvipsnames]{xcolor}
\usepackage[caption=false]{subfig}
\usepackage[super]{nth}
\usepackage{array}
\usepackage{multirow}
\usepackage{subfiles}

\allowdisplaybreaks

\usepackage[normalem]{ulem}

\def\be{\begin{equation}}
\def\ee{\end{equation}}
\def\ba{\begin{eqnarray}}
\def\ea{\end{eqnarray}}

\newcommand{\yes}{$\checkmark$}
\newcommand{\no}{$\times$}
\newcommand{\maybe}{$\bigtriangleup$}
\newcommand{\na}{$-$}

\begin{document}
\title{Observation of broken inversion and chiral symmetries in the pseudogap phase in single and double layer bismuth-based cuprates}

\author{Sejoon Lim}
\email{lims@stanford.edu}
\affiliation{Department of Applied Physics, Stanford University, Stanford, California 94305, USA}
\affiliation{Stanford Institute for Materials and Energy Sciences, SLAC National Accelerator Laboratory, 2575 Sand Hill Road, Menlo Park, California 94025, USA}
\author{Chandra M. Varma}
\thanks{Recalled professor}
\affiliation{Physics Department, University of California, Berkeley, California 94704}
\author{Hiroshi Eisaki}
\affiliation{National Institute of Advanced Industrial Science and Technology, Tsukuba, Ibaraki 305-8568, Japan}
\author{Aharon Kapitulnik}
\affiliation{Department of Applied Physics, Stanford University, Stanford, California 94305, USA}
\affiliation{Stanford Institute for Materials and Energy Sciences, SLAC National Accelerator Laboratory, 2575 Sand Hill Road, Menlo Park, California 94025, USA}
\affiliation{Department of Physics, Stanford University, Stanford, California 94305, USA}

\begin{abstract}
We deduce the symmetry of the pseudogap state in the single and double layer bismuth-based cuprate superconductors by measuring and analyzing their circular and linear photogalvanic responses, which are related {\it linearly} to the chirality and inversion breaking respectively of the order parameter. After separating out the trivial contribution arising from the surface where inversion symmetry is already broken, we show that both responses start  below the pseudogap temperature $T^*$ and grow below it to a sizable magnitude, revealing the broken symmetries in the bulk of the crystal. Through a detailed analysis of the dependence of the signals on the angle of incidence, the polarization of the light, and the orientation of the crystal, 
 we are able to discover that the point group symmetry below $T^*$ is limited to $mm2$ or $mm2\underline{1}$ groups. Taking into account formation of domains and previous measurements, our results narrow down the possible symmetries of the microscopic origin of the phase transition(s) at $T^*$. 
 \end{abstract}

\date{\today}
\maketitle

\date{\today}
\maketitle
\section{Introduction}

One of the long standing questions in the study of the high-Tc cuprate superconductors is the understanding of the nature of the boundary between the strange metal phase and the pseudogap phase \cite{keimer15}. Over the years, it has become increasingly evident through experimental findings, that this boundary marks a true phase transition, initially suggested in \cite{varma97} proposing the so-called loop-order model, with subsequent range of models exhibiting magnetic (e.g. \cite{Chakravarty2001}) or charge order (for a review see e.g. \cite{Fradkin2015}). Some of the observed broken symmetries in the canonical yttrium- and bismuth-based cuprates, for example, include time-reversal \cite{fauque06,xia08,mook08,kaminski02,he11,dealmeidadidry12,manginthro14,zhao17}, four-fold rotation \cite{daou10,comin15,sato17,howald03,Vershinin2004,Wise2008,lawler10,parker10,dasilvaneto14,Mukhopadhyay2019,ishida20}, translation \cite{comin15,howald03,Vershinin2004,Wise2008,parker10,dasilvaneto14,Peng2016,Mukhopadhyay2019}, inversion \cite{zhao17}, or a combination of these \cite{fauque06,mook08,kaminski02,he11,dealmeidadidry12,manginthro14,zhao17}.
However, while all those experiments contribute to constrain the symmetry  of the pseudogap state, it is not yet considered fully understood. In particular, the question of interplay between structural and electronic effects and the generality of the resulting pseudogap symmetry among the different cuprates needs to be sharpened up. For example, while short range charge order phase appears well separated from the onset of pseudogap in underdoped YBCO cuprates \cite{keimer15}, the two are found to coincide in the BSCCO cuprates (e.g. \cite{he11,dasilvaneto14,Mukhopadhyay2019}). 

In this paper, we report circular (CPGE) and  linear (LPGE) photogalvanic effects (PGE) \cite{belinicher80,sturman92,ganichev03} in single crystals of near optimally doped single and double layer bismuth-based cuprate superconductors (BSCCO): Pb\textsubscript{0.55}Bi\textsubscript{1.5}Sr\textsubscript{1.6}La\textsubscript{0.4}CuO\textsubscript{6+$\delta$} (Pb-Bi2201) with $T_c \approx 38$ K and Bi\textsubscript{2.1}Sr\textsubscript{1.9}CaCu\textsubscript{2}O\textsubscript{8+$\delta$} (Bi2212) with $T_c \approx 86$ K, previously used in studies in our group \cite{he11,Fang2004}. While PGE originating from the ``trivial'' inversion symmetry breaking of the probed surface is present already at temperatures above $T^*$, an additional signal with onset near the pseudogap temperature $T^*$ and growing with decreasing temperature is clearly discerned. Here we present evidence that the appearance of this ``nontrivial'' PGE across $T^*$ points to inversion symmetry breaking and chirality in the pseudogap state, which is robust to spatial averaging of structural domains. The CPGE and LPGE observed are linear in the photon intensity and therefore linear in the order parameter characterizing the broken symmetry of the pseudogap phase, unlike almost all the scattering experiments mentioned above. This allows us through a comprehensive symmetry analysis of the experiments to deduce more details of the symmetry than other techniques which have been used to date.

The PGE is the dc current measured in response to photon intensity with specified polarization and angle of incidence:
\begin{equation}
j_i =\chi_{ijk} (E_j E_k^* + E_j^* E_k )/2 + i \gamma_{il} (\vec{E} \times \vec{E}^* )_l, 
 \label{pge}
\end{equation}
where $\vec{E}$ is the complex amplitude of the electric field with components $\{E_i\}$ and intensity $I=|\vec{E}|^2$. The CPGE is characterized by the second-rank axial tensor $\gamma_{il}$, and the LPGE by the third-rank polar tensor $\chi_{ijk}$ \cite{birssnotations}. 
Both effects are absent in centrosymmetric media. In Eq.~(\ref{pge}), effects due to  photon drag and sample heating are not included.

The summary of our observations  of the nontrivial PGE is as follows: \textit{i}) We observe chirality in both CPGE and LPGE, i.e. the direction of current in the plane is rotated with respect to the plane of incidence and its sign is invariant to the in-plane rotation of the crystal; \textit{ii}) Both the in-plane CPGE and LPGE are observed only at oblique incidence of radiation; \textit{iii}) At normal incidence, only the $c$-axis LPGE is observed; \textit{iv}) The magnitude of the PGE is of the order of the surface-induced PGE for Bi2212 and several times larger for Pb-Bi2201. 

These observations and the detailed dependence of $j_i$ on the angle of the plane of  incidence of the photons and their polarization provide 11 different criteria which are used to discover the point group symmetry of the pseudogap phase after an examination of the predictions for all the relevant 34 groups. These groups and their consistency with each of the 11 properties or otherwise are listed  in Table S1 in the Supplemental Material \cite{supplemental}.

\section{Experiment}

\begin{figure}[ht]
\includegraphics[width=0.95\linewidth]{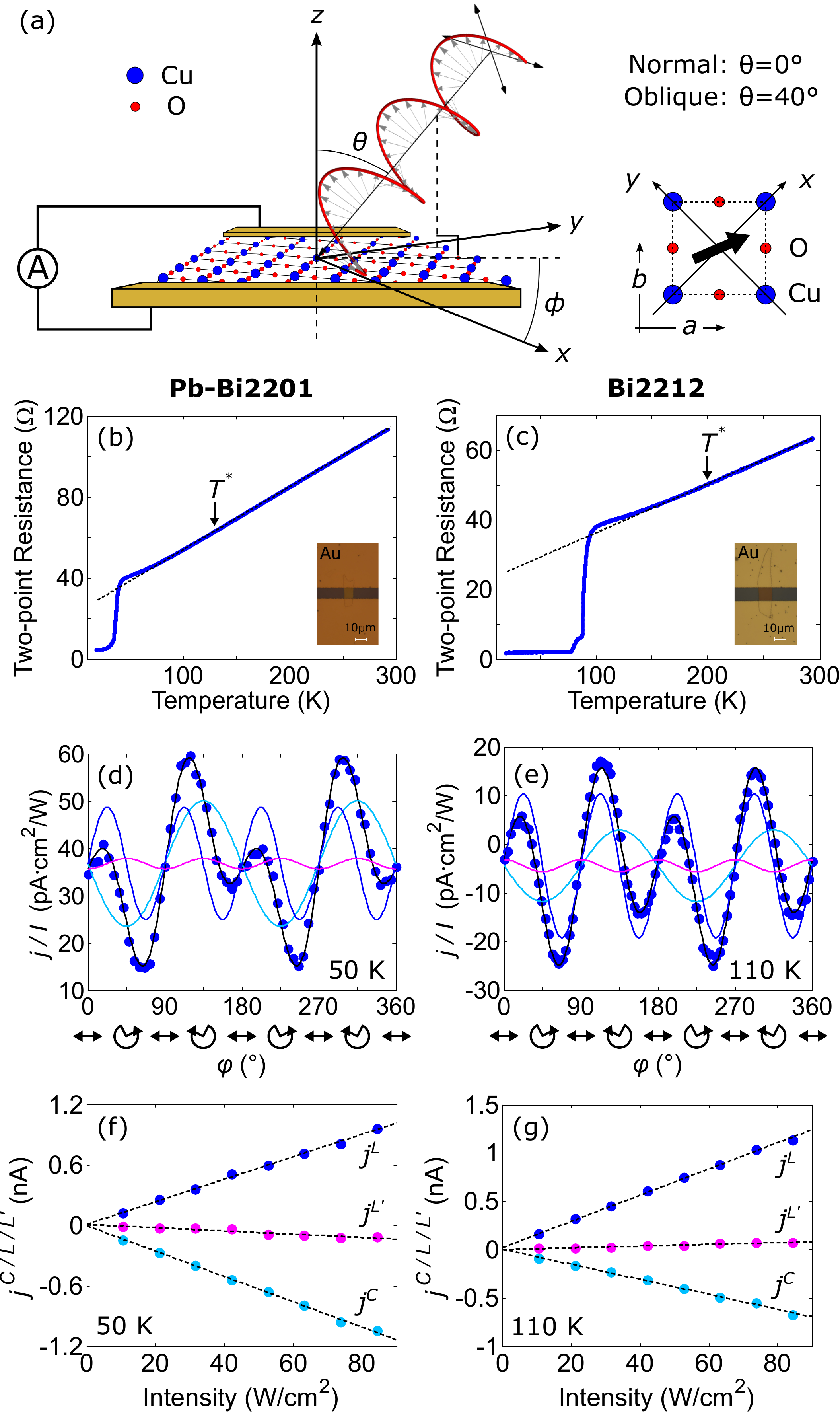}
\caption{Measurements of the in-plane electrical resistance and photoinduced electric current on a 70 nm thick Pb-Bi2201 device and a 100 nm thick Bi2212 device. (a) Schematic of the experimental geometry. (b,c) Two-point electrical resistance of (b) Pb-Bi2201 and (c) Bi2212 as a function of temperature. The insets show optical images of the measured devices. (d,e) Photoinduced electric current measured at (d) 50 K in Pb-Bi2201 and at (e) 110 K in Bi2212 as a function of the phase angle $\varphi$. The measurements were carried out under oblique incidence of radiation. In each plot, the black line shows a fit to the phenomenological equation, while the light blue, blue, and magenta lines show the components proportional to $\sin 2\varphi$, $\sin 4\varphi$, and $\cos 4\varphi$, respectively. (f,g) Intensity dependence of $j^{C}$, $j^{L}$, and $j^{L'}$ in (f) Pb-Bi2201 and (g) Bi2212.}
\label{fig1}
\end{figure}
Two-terminal BSCCO devices were fabricated as described in the Supplemental Material \cite{supplemental}, and mounted on the PGE apparatus as illustrated in Fig.~\ref{fig1}(a). Electrical resistance as a function of temperature was first measured to characterize the samples and obtain an estimate of $T^*$ by locating the point of deviation from linear resistivity (see e.g.\ \cite{Sterpetti2017}). Typical resistance curves for Pb-Bi2201 with $T_c \approx 35$ K and $T^* \approx 130$ K and Bi2212 with $T_c \approx 87$ K and $T^* \approx 200$ K are shown in Fig.~\ref{fig1}(b) and Fig.~\ref{fig1}(c), respectively.

PGE measurements were carried out for a fixed angle $\theta$ at either normal ($\theta = 0^\circ$) or oblique ($\theta = 40^\circ$) incidence of radiation at 1550 nm but at several values of the azimuthal angle $\phi$. The photoinduced electric currents were measured as a function of the phase angle $\varphi$, which is the rotation angle between the plane of the initial \textit{p}-polarization and the optical axis of a quarter-wave plate. Emphasizing CPGE, we use a quarter-wave plate, rotating it continuously to scan between right and left circular polarizations. Thus, in our experiment information about LPGE is extracted from the intermediate polarizations that vary between pure linear in the incidence plane, to elliptical at an arbitrary angle. 

Figures~\ref{fig1}(d) and \ref{fig1}(e) show typical measurements carried out under oblique incidence at temperatures near but above $T_c$. Overall, these currents share a fairly similar dependence on the polarization of the excitation light, and are well fitted to the phenomenological equation \cite{mciver12}
\be
j = j^{C} \sin 2\varphi + j^{L} \sin 4\varphi + j^{L^\prime} \cos 4\varphi + d
\label{fit}
\ee
Here $j^{C}$ is the CPGE coefficient, while both $j^{L}$ and $j^{L^\prime}$ contain information about LPGE. The fits of the data to the above equation exhibit the following common features: \textit{i}) $j^{C}$, $j^{L}$, and $j^{L'}$ are proportional to the light intensity as shown in Fig.~\ref{fig1}(f) for Pb-Bi2201 and Fig.~\ref{fig1}(g) for Bi2212;  \textit{ii}) Where observed at oblique incidence, $j^{C}/j^{L}\sim\mathcal{O}(1)$ and they follow the same temperature dependence but with opposite sign; \textit{iii}) For $y$-$z$ or $x$-$z$ incident planes (i.e. current measured along $x$ or $y$ respectively), $j^{L^\prime}$ is typically small, on the order of $\sim0.1j^L$; \textit{iv}) For incident plane rotated away from the principal axes ($x$ or $y$), $j^{L^\prime}$ becomes finite beyond its typical residuals, and may increase in magnitude to be a large fraction of  $j^{L}$. We will use this information to determine that  true inversion symmetry breaking in the material is manifested by the CPGE term $j^{C}$, and by the LPGE terms $j^{L}$ and $j^{L^\prime}$ (when the two terms are of same order). The constant term $d$, and the residuals of $j^{L^\prime}$ are associated with the photon drag and thermal effects (see e.g.\cite{ sturman92}). (see further discussion in the Supplemental Material \cite{supplemental}).

\section{Results}

\begin{figure}[ht]
\includegraphics[width=1.0\linewidth]{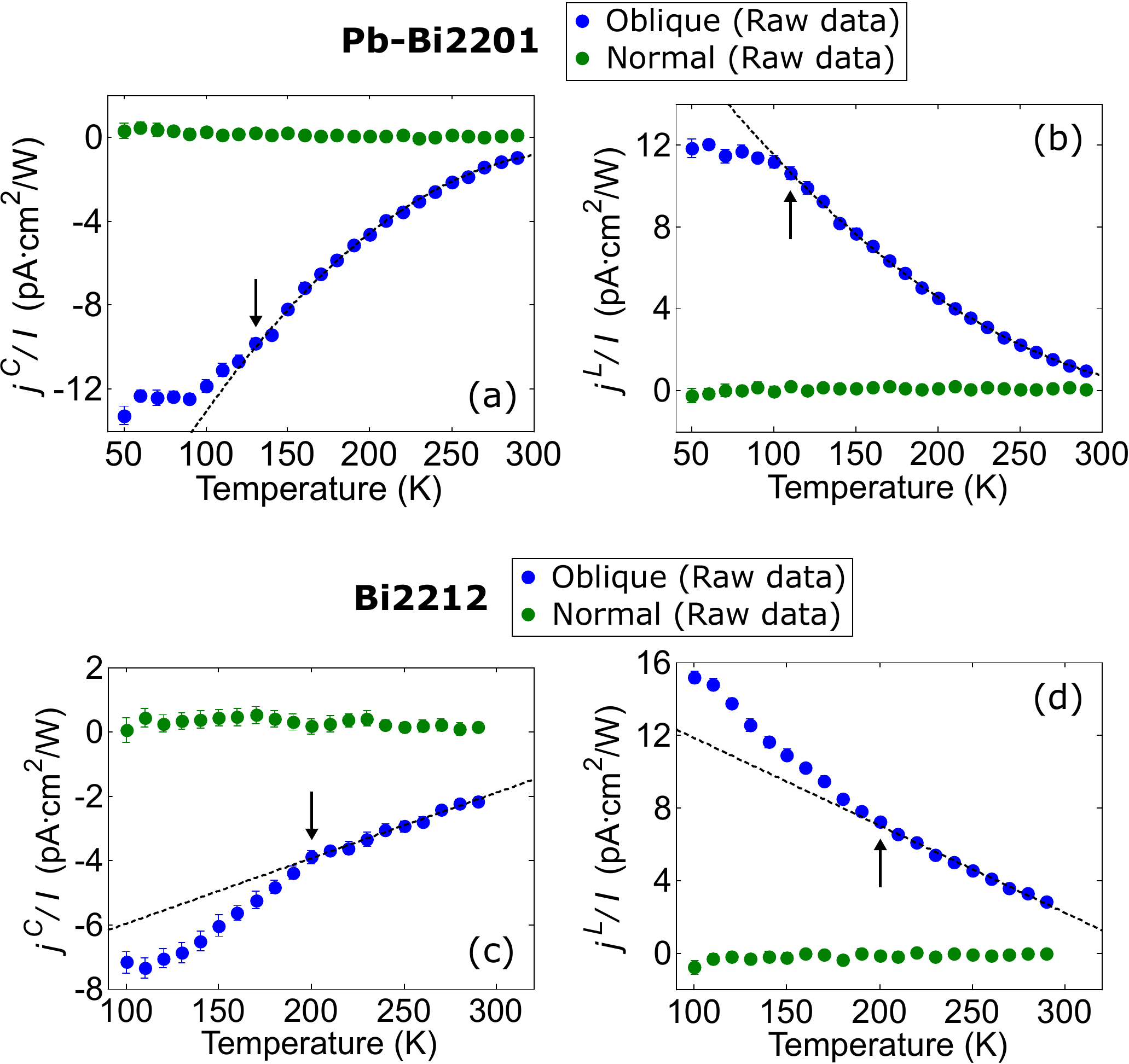}
\caption{Temperature dependence of $j^C$ and $j^L$, normalized by the light intensity $I$, in the Pb-Bi2201 and Bi2212 samples characterized in Fig.~\ref{fig1}. The dashed black lines are guides to the eye, and the arrows mark the approximate onset of deviation from the high-temperature trend.}
\label{fig2}
\end{figure}
Figure~\ref{fig2} shows the temperature dependence of $j^{C}$ and $j^{L}$, normalized by the light intensity $I$, in the Pb-Bi2201 and Bi2212 samples characterized in Fig.~\ref{fig1}. In Pb-Bi2201, where Pb doping suppresses the superstructure in the BiO plane \cite{he11}, the current direction is estimated to be along one of the principal axes, $x$ or $y$. The current direction in Bi2212 is carefully aligned close to the $y$-axis in the direction of the superstructural modulation \cite{kirk88} (see Supplemental Material \cite{supplemental}). 

A feature that complicates the analysis of the oblique incidence data is the occurrence of $j^C$ and $j^L$ already at room temperature. Starting from a bulk orthorhombic symmetry consistent with point group symmetry $mmm$ \cite{Miles1998,Mans2006,DiMatteo2007}, with the surface also orthorhombic \cite{Mans2006}, we do not expect PGE from this centrosymmetric bulk. However, for a simple surface characterized by a single surface-normal vector, both CPGE and LPGE are allowed at oblique incidence while only the LPGE is allowed  at normal incidence (see Supplemental Material \cite{supplemental}). Moreover, a monotonic and featureless variation of this surface contribution with decreasing temperature is expected from a simple kinetic approach, where the increasing mean free path results in the increasing magnitude of PGE currents \cite{Deyo2009}. We thus treat this surface contribution as a trivial baseline, relative to which nontrivial contributions are observed.

Indeed, with  decreasing temperature, $j^C$ and $j^L$ start to deviate from their high-temperature trend near $T^*$, indicating the appearance of nontrivial PGE currents.  These relative deviations are much larger than the subtle deviations of resistivity through $T^*$ and cannot be explained by the kinetic approach  \cite{Deyo2009}. This behavior is more pronounced in Pb-Bi2201, where the rate at which $j^C$ and $j^L$ increase with decreasing temperature slows down below $\sim$130 K. In Bi2212, a similar deviation, but with opposite polarity, is observed roughly near 200 K, which is somewhat weaker in magnitude and sets in gradually over a broader temperature range. 

\begin{figure}[ht]
\includegraphics[width=1.0\linewidth]{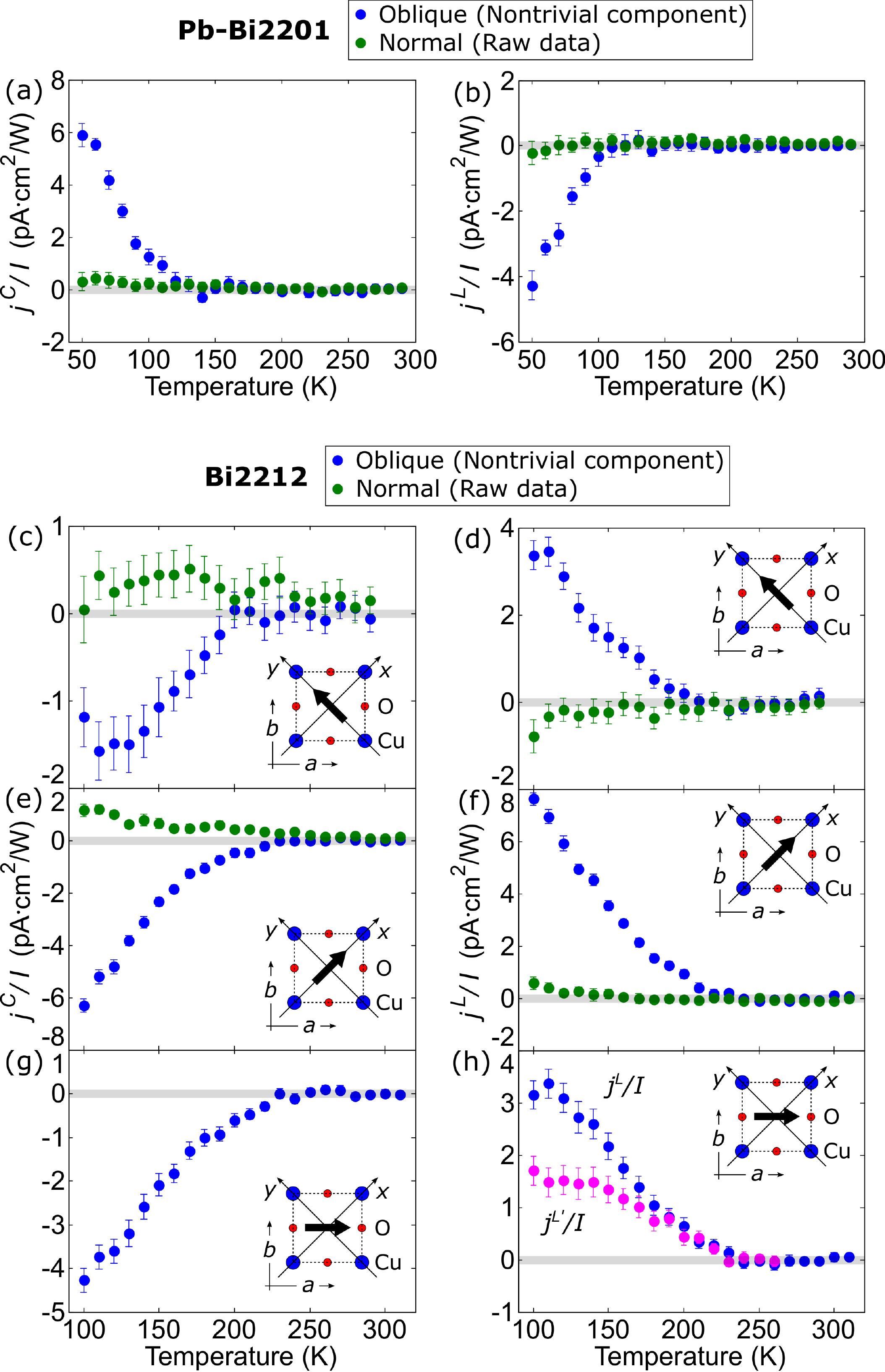}
\caption{Temperature dependence of $j^C/I$ and $j^{L}/I$, where the oblique incidence data has been normalized to show the nontrivial component of the overall current.  (a,b) Data for Pb-Bi2201 with current direction estimated along $x$ or $y$.  (c-h)  Data for Bi2212 in three different configurations. Current direction aligns close to either the (c,d) $y$-axis, (e,f) $x$-axis, or (g,h) Cu-O bond direction. The magenta data points in  (h) represent $j^{L'}/I$ in this configuration. This LPGE term is small (hence, not displayed), in the other two configurations shown in (d) and (f) (see Supplemental Material \cite{supplemental}). }
\label{fig3}
\end{figure}

Fitting the high temperature oblique incidence PGE data to a smooth curve, we ``normalize'' the data by subtracting this trivial background from the overall current to extract the low-temperature evolution of the nontrivial component. Figure \ref{fig3}(a-b) show the nontrivial CPGE and LPGE vs. temperature for Bi2201, suggesting a behavior consistent with an increasing order parameter below $T^*$. Figure~\ref{fig3}(c-h) summarizes the more comprehensive study carried out on Bi2212, where three different samples from the same batch are carefully aligned with current directions close to either the $y$-axis (c,d), $x$-axis (e,f), or Cu-O bond direction (g,h) (see Supplemental Material \cite{supplemental}). Note that normal incidence data in Fig.~\ref{fig3} is (unsubtracted) raw data, which seems to agree with no PGE in that configuration for either material.

Focusing on Bi2212, a key feature of all these measurements is that the nontrivial PGE currents show similar size and trend  irrespective of the crystal orientation, which suggests a chiral behavior. However, cooling the sample in the presence of intense light with either right or left circular polarization prior to the PGE measurements did not yield different results, thus suggesting that no simple trainable gyrotropic effect is present (see e.g. \cite{Xu2020}). We are therefore led to include magnetic symmetry groups in analyzing the CPGE. 

\begin{figure}
\includegraphics[width=1.0\linewidth]{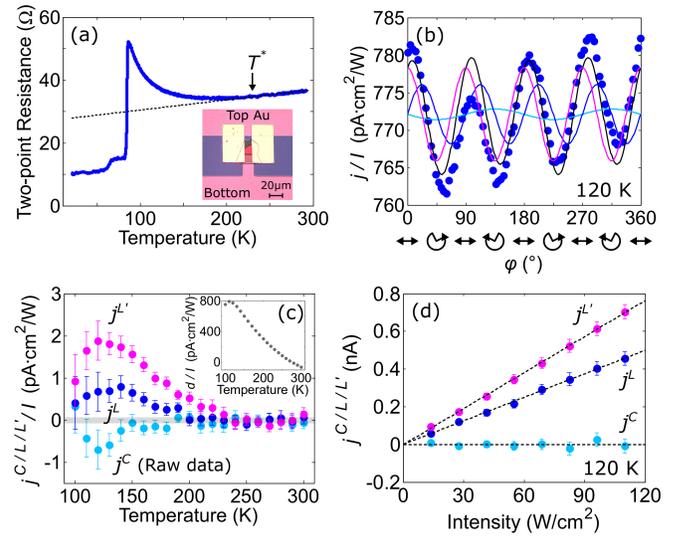}
\caption{Measurements of the $c$-axis electrical resistance and photoinduced electric current on a 90 nm thick Bi2212 device. (a) Two-point electrical resistance as a function of temperature. The inset shows an optical image of the measured device. (b) Photoinduced electric current measured at 120 K as a function of the phase angle $\varphi$. The measurement was carried out under normal incidence of radiation. The black line shows a fit to the phenomenological equation, while the light blue, blue, and magenta lines show the components proportional to $\sin 2\varphi$, $\sin 4\varphi$, and $\cos 4\varphi$, respectively. (c) Temperature dependence of $j^C$, $j^L$, and $j^{L'}$, normalized by the light intensity $I$. $j^L$ and $j^{L'}$ have been normalized to show the nontrivial component. The inset shows the temperature dependence of $d/I$. (d) Intensity dependence of $j^C$, $j^L$, and $j^{L'}$.}
\label{fig4}
\end{figure}

To complement the $a$-$b$ plane current measurements, we also searched for  PGE along the $c$-axis in Bi2212. Figure \ref{fig4} is an example of such a measurement, where Fig.~\ref{fig4}(a) shows a typical $c$-axis resistance curve with $T^*\approx230$ K marked as the point of deviation from high-temperature linear resistivity. The $c$-axis photocurrent measured under normal incidence of radiation at 120 K is shown in Fig.~\ref{fig4}(b). Although there is a significant offset component arising from the photon drag and thermal effects, as well as an accompanying slight distortion in the overall waveform due to alignment imperfections, we can extract  $j^C$, $j^L$, and $j^{L'}$ by following the fitting procedure of Eq.~(\ref{fit}). Figures~\ref{fig4}(c) and~\ref{fig4}(d) depict  the temperature and light intensity dependence of these fitting coefficients, indicating that, while small, only the LPGE occurs in this measurement geometry. The LPGE surface term above $T^*$ is of the same order of magnitude and similar in trend to the $a$-$b$ plane LPGE  \cite{supplemental}, reflecting the similar decrease in sample $c$-axis resistance in that regime. Thus, the sharp increase in $c$-axis resistance below $T^*$ would yield a large decrease in surface LPGE in that direction, while we observe the opposite trend, confirming a nontrivial $c$-axis LPGE below $T^*$.

\section{Analysis and Discussion}

\subsection{Symmetry analysis}

As discussed above and in SM \cite{supplemental}, surface effects cannot explain the emergence of PGE deviations below $T^*$. We therefore turn to the possible broken symmetries in the bulk  in the pseudogap regime, which yield PGE. While LPGE calls only for broken inversion symmetry, the observation of chirality in both LPGE and CPGE, and the fact that time-reversal symmetry breaking has been reported for both Pb-Bi2201 \cite{he11} and Bi2212 \mbox{\cite{kaminski02,dealmeidadidry12,manginthro14}}, suggests that we search among the  magnetic point groups including the allowed classical subgroups \cite{birss64}. In carrying out symmetry analysis, we start with the general requirements for $\gamma_{il}$ \cite{Deyo2009} and identify the possible magnetic point groups consistent with the CPGE data. Including ``gray groups'' (see below), this procedure yields 34 point groups in the triclinic, monoclinic, orthorhombic and tetragonal groups, which we need to test in accord with the observations \textit{i}), \textit{ii}) and \textit{iii}) which extend to 11 properties when we take into account the angular and polarization dependences described above. All 34 point groups are listed in Table S1 \cite{supplemental}, together with the 11 properties and marked with the consistency or inconsistency in each group for each of the properties.  In what follows we use the notations where $\underline{o}$ denotes a regular operator $o$ combined with time reversal symmetry operator, while $\bar{n}$ denotes a $n$-fold rotation-inversion operator.
  
We first examine point groups allowed by CPGE where the principal axis lies along the $c$-axis (i.e. $z$-axis).  It is common to approximate the crystal structure of  BSCCO as tetragonal (since orthorhombic distortions mostly affect the BiO planes with almost no effect on the CuO planes \cite{Miles1998,Zeljkovic2012}), for which the 4-fold symmetry groups allowed are $\underline{4}m\underline{m}$ and $\underline{4}$. However, the true bulk crystal of both, Bi2201 or Bi2212 is orthorhombic above $T^*$ \cite{{Miles1998,Mans2006,DiMatteo2007}}, with a slight distortion that reduce the symmetry around the $y$-axis, but still remain orthorhombic \cite{Gladyshevskii1996}.  Thus, if we take into account this orthorhombic distortion, then $mm2\underline{1}$ and any of its subgroups ($mm2$,~$\underline{mm}2$,~$\underline{m}m\underline{2},~$m\underline{1}$,~$\underline{m}$, ~$m$,~$2\underline{1}$,~$\underline{2},~$2$, $1\underline{1}$ and $1$), are also allowed, where the monoclinic subgroups may require a different principal axis (see table S1 in SM \cite{supplemental}).

We next use the LPGE data to narrow down the list of groups by  searching for  $\chi_{ijk}$ tensors that match each of the properties of the LPGE data. Here, in addition to the requirement of chirality, a key feature in the LPGE Bi2212 data is that the coefficient $j^{L^\prime}$ is vanishingly small along $x$ or $y$, but $j^{L^\prime}\sim j^L$ along the Cu-O direction (e.g.,  below $T^*$,  $j^{L^\prime} \approx 0.5 j^L$, see Fig.~\ref{fig3}(h)). Implementing these two effects,  and noting that the in-plane symmetry is nearly 4-fold, which suggest similar magnitude to similar tensor components, both which are discussed at length in the Supplemental Material \cite{supplemental}, we find that only the sub-groups $mm2\underline{1}$ and $mm2$ are fully consistent with both our CPGE and LPGE data. 

However, we note that PGE is proportional to the order parameter, and thus the PGE current that appears upon uniform illumination of the sample is very sensitive to mesoscopic domains which if random, would average the PGE current to $\mathcal{O}\big((d/L)^2\big)$, where $d$ is the typical size of the domains and $L\sim 10\mu$m is the size of the illuminated sample. This issue is further discussed next.

\subsection{The results in a broader context}

Focusing on the above result, $mm2$  has the following symmetries, $1, \overline{2}_x, \overline{2}_y, 2_z$, respectively, identity, two -fold improper rotation about the x and y- axes, and two-fold proper rotation about the z-axis. $mm2\underline{1}$ is the equivalent ``gray group''  (see e.g. \cite{Lifshitz2005}), where $\underline{1}$ represents the addition of time reversal operator to the group, which effectively implies that for each moment in the unit cell there is the opposite moment at the same position. Thus, a ``gray group'' is often used to describe an equivalent paramagnetic state within the same crystallographic group, which is invariant under time reversal.  A cartoon for the smallest orthorhombic unit-cell exhibiting the symmetries $mm2$ and $mm2\underline{1}$ is shown in the SM \cite{supplemental}.

On the face of it, our analysis yields an orthorhombic symmetry for the pseudogap state, which can also be taken as non-magnetic in origin.  For example, since BSCCO above $T^*$  is orthorhombic with $Bb2b$ space group due to a slight distortion around the $y$-axis \cite{Kan1992} (thus, equivalent to $mm2$ with rotation around $y$ \cite{DiMatteo2007}), an electronic driven structural effect that sets in at $T^*$, such as a charge order or nematic transition \cite{kivelson98,kivelson03,vojta09,fradkin10,lawler10}, would reduce the initial orthorhombic symmetry to monoclinic (note that charge order in BSCCO appears along the Cu-O bonds, thus at 45$^\circ$ to the principal axes). Small enough monoclinic domains can then assemble to yield an effective $mm2$ symmetry with two-fold rotation along the $c$-axis. However, PGE is proportional to the order parameter, and thus the PGE current that appears upon uniform illumination of the sample is very sensitive to averaging of mesoscopic domains. For random domains that carry opposite sign of the order parameter, PGE current will be reduced by a factor of $\mathcal{O}\big((d/L)^2\big)$, with a standard-deviation $\mathcal{O}(d/L)$, where $d$ is the typical size of the domains and $L\sim 10\mu$m is the size of the illuminated sample.  For BSCCO system, a one-dimensional charge density wave with typical domain size of $\sim$30 -100 \AA~ has been consistently observed in both Pb-Bi2201 \cite{Wise2008,he11} and Bi2212 \cite{howald03,Vershinin2004,parker10,dasilvaneto14,Mukhopadhyay2019} below the pseudogap state. Such small domains would predict a $\sim 10^{-4}$~to~$10^{-3}$ and often much larger reduction of the single domain signal, which would make the PGE signal impossible to observe. We note that compared to ``standard'' materials, typical free carrier CPGE in tellurium \cite{Asnin1979,Tsirkin2018}, or LPGE in heavily doped GaAs \cite{Andrianov1982} are at most a factor of 10 larger than our nontrivial PGE values. 

Thus, any smaller, monoclinic domains that assemble to exhibit $mm2$ at the mesoscopic scale must be of a certain type as to avoid a reduction of PGE currents inside a domain. In our system the key issue will be to maintain the intrinsic chirality of the domains at the sample scale. Solving this issue will also help to understand previous observation of  X-ray natural circular dichroism (XNCD) that appears below $T^*$ \cite{Kubota2006}, and was argued to demonstrate that time reversal symmetry is preserved in the pseudogap phase of underdoped Bi2212. These results were initially demonstrated to be inconsistent with only a crystal structure effect, without TRSB \cite{DiMatteo2007}. However, a monoclinic distortion and domain structure that preserves chirality, which is needed to explain the PGE results, can now explain the observed XNCD.  Thus, simultaneous TRSB effects cannot be ruled out since the geometry of the experiment, with X-rays wavevector in the $c$-direction may not be sensitive to magnetic effects with in-plane order parameter, that in BSCCO appear concurrent with charge order at $T^*$ \cite{kaminski02,dealmeidadidry12}.

On the other hand, a solely structural effect, even with the addition of charge order transition, without TRSB is at odds with other experiments that specifically probe time reversal and inversion symmetry breaking. In particular, if we include magnetic moments (either spins, or current-loops), unless moments are intra unit-cell, $mm2$ structure  breaks translational symmetry, which is inconsistent with neutron diffraction experiments.  The polarized magnetic scattering \cite{fauque06,mook08,Li2008} including Bi2212  \cite{dealmeidadidry12} observe extra intensity below $T^*$ at the $(1,0,{\ell})$ Bragg spots (and its equivalents due to domains) consistent with ${m\underline{m}m}$ (with twofold axis rotation around $y$ \cite{simon03}), and nothing at the much easier to observe $(1/2,1/2, \ell)$ Bragg spots, which would be required for $mm2$ of any origin.

Second-harmonic generation (SHG) was also suggested \cite{simon03} as an effective probe for bulk inversion symmetry breaking, and performed on YBa$_2$Cu$_3$O$_{6+x}$ (YBCO) \cite{zhao17}.  The data over a wide doping range revealed a monoclinic crystal with symmetry $2/m$ (only two-fold proper rotation about the c-axis and a mirror plane, rather than orthorhombic already above $T^*$ (assumed to be due to disorder in the oxygen chains). Below $T^*$, the data is consistent with domains of intrinsic symmetry $\underline{2}/ m$, or  $m\underline{1}$. The observed SHG was then interpreted as an incoherent response from domains, smaller than the laser spot, that average out to maintain an observed C$_2$ symmetry below $T^*$ \cite{zhao17}. Unlike BSCCO, in YBCO the charge order onsets below a characteristic temperature $T_{CO} < T^*$, and resembles more a crossover than a true order parameter, presumably partially due to disorder (see e.g. phase diagram in \cite{keimer15}). Thus, the expectation that the pseudogap is a universal phenomenon within the cuprates suggests that the transition that we observe in BSCCO at $T^*$ will have similar origin to the sharp onset of order in SHG experiments on YBCO. This further implies that charge order alone cannot be the only explanation to the observed symmetry deduced from the PGE data below $T^*$. For example, $m\underline{1}$ observed in SHG in YBCO is a subgroup of  $mm2\underline{1}$ observed in PGE in BSCCO, which could point to a similar origin of the pseudogap order parameter in the two materials. 

Thus, the above discussion suggests that  to understand the PGE results we need to consider mesoscopic domains of lower symmetry that when fused together continue to satisfy the required constraints from the data, particularly the chiral behavior. Assuming the same symmetry breaking as that observed in SHG, our data will be consistent with domains of $m\underline{1}$ rotated 90$^\circ$, thus averaged out to yield the observed $mm2\underline{1}$ symmetry, while also allow for an order parameter that is odd under time reversal. The lack of mirror symmetries along the $c$-direction reflects the observed chirality, which could be unique to the BSCCO system, e.g. associated with distortions in the Bi-O layer \cite{Gladyshevskii1996}. As it is constrained to the Bi-O layers, it may not interfere with the intra-unit cell loop order observed in neutron scattering. 

Several different models exhibiting TRSB  were proposed to explain the symmetry breaking below $T^*$ \cite{varma97,simon02,varma06,yakovenko2015,Lovesey2015,Fechner2016}. However, to agree with our observed  $mm2\underline{1}$ or $mm2$ symmetry, they must impose ``domain-fusing,'' which maintains coherence over the size of the sample to account for the magnitude of the effect we observe. As we discussed above, this may not be a simple task, since in general domain averaging tends to reduce a signal.  A model that respects our observed symmetry and relies on topology to guarantee that  PGE currents within domains add coherently was recently proposed in \cite{varma2019}. In that model, the four possible domains with internal current loop order with unit-cell $m\underline{m}m$ symmetry \cite{simon02} observed in neutron scattering \cite{dealmeidadidry12} are glued together subject to the requirement that currents at the domain boundaries are conserved leading to a unique super-cell. The resulting boundary currents from domains of size of  5 or more lattice constants (see Fig.~S5(c) in \cite{supplemental}) respect our observed $mm2$ symmetry. Further discussion of this possibility is given in the Supplemental Material \cite{supplemental}.
 
Finally, with a magnetic origin for both, $mm2\underline{1}$ and $mm2$, which equally well agree with the data. The former is the paramagnetic version of the latter, which would result if the moments were anisotropic and fluctuating in a time-scale faster than the time-scale of our measurements (or if instead of ordered moments there were ordered quadrupoles.) This is reminiscent of the contrast between neutron scattering with a time-scale of measurement smaller than $10^{-12}$ secs over which order is observed \cite{fauque06,mook08,dealmeidadidry12,manginthro14} and $\mu$-relaxation rate. The latter, which is also linear in the order parameter, observes an altered rate below $T^*$, from which an internal magnetic field below $T^*$ fluctuating at a time scale of $\mathcal{O}(10^{-7})$ secs is deduced \cite{Shu2018,gheidi20}. Our measurements are basically dc and would therefore be consistent with these results.

\section{Summary}

In summary, through detailed observation of circular and linear photogalvanic effects we show that the transition to the pseudogap regime in two families of Bi-based cuprate superconductors marks a phase transition associated with the development of chiral and inversion symmetry breaking. Our results are shown to be consistent with previous neutron scattering results, while also explain the previously puzzling observation of the onset of chirality below $T^*$ \cite{Kubota2006}.  While charge order, which for the BSCCO system was shown to occur at $T^*$  and thus could potentially lower the crystal symmetry from orthorhombic to monoclinic, domain averaging, will retain the robust chiral and inversion symmetry breaking, while exhibiting an effective domain symmetry consistent with the magnetic point groups $mm2\underline{1}$ and $mm2$. At the same time, searching for a common mechanism for the pseudogap in all cuprates, including YBCO, we must conclude that charge order transition in BSCCO may affect the observed symmetry, but is not the main cause of the pseudogap order parameter. We finally note that certain models of intra unit cell loop current order can reassemble into domains which is consistent with $mm2\underline{1}$ and $mm2$ symmetries, where the former can explain the lack of observations local magnetism at low frequencies, while the full intra-unit-cell antiferromagnetic loop order is observed at short (i.e. neutron scattering) time scales.

\section*{Acknowledgements}
The authors thank Yoshiyuki Yoshida for help with growing single crystals of Pb-Bi2201 previously characterized in \cite{he11}, and Nobuhisa Kaneko and Martin Greven for providing us single crystals of Bi2212, previously characterized in \cite{Fang2004}. Stimulating discussions with B. Spivak, D. Hsieh, S. Kivelson, T. Morimoto and M. Norman are greatly appreciated. This work was supported by the U.S.~Department of Energy, Office of Science, Basic Energy Sciences, Division of Materials Sciences and Engineering, under Contract DE-AC02-76SF00515. Device fabrication was partially supported by the Gordon and Betty Moore Foundation through Grant GBMF4529.

\bibliography{Ref}

\begin{thebibliography}{68}
\bibitem [{bir()}]{birssnotations}%
  \BibitemOpen
  \href@noop {} {}\bibinfo {note} {Using notations of Birss \cite{birss64} in
  considering effect of time-inversion, we identify CPGE with an ``i''-tensor
  and LPGE with a ``c''-tensor}\BibitemShut {NoStop}%
\end{thebibliography}

\newpage

\onecolumngrid
\newpage
\setcounter{section}{0}
\setcounter{figure}{0}
\renewcommand{\thefigure}{S\arabic{figure}}
\renewcommand{\theequation}{S.\arabic{equation}}
\renewcommand{\thetable}{S\arabic{table}}
\renewcommand{\thesection}{S\arabic{section}}

\renewcommand{\thefootnote}{\fnsymbol{footnote}}

\begin{center}
\textbf{ SUPPLEMENTARY INFORMATION}

\vspace{3em}
\textbf{Observation of broken inversion and chiral symmetries in the pseudogap phase in single and double layer bismuth-based cuprates}\\

\fontsize{9}{12}\selectfont

\vspace{3em}
Sejoon Lim,$^{1,2}\footnote{ lims@stanford.edu}$~Chandra Varma,$^{3}\footnote{ Recalled Professor}$~Hiroshi Eisaki,$^4$ and Aharon Kapitulnik$^{1,2,5}$\\

\vspace{1em}
$^1${\it Department of Applied Physics, Stanford University, Stanford, California 94305, USA}\\
$^2${\it Stanford Institute for Materials and Energy Sciences, SLAC National Accelerator Laboratory, 2575 Sand Hill Road, Menlo Park, California 94025, USA}\\
$^3${\it Physics Department, University of California, Berkeley, California 94704}\\
$^4${\it National Institute of Advanced Industrial Science and Technology, Tsukuba, Ibaraki 305-8568, Japan}\\
$^5${\it Department of Physics, Stanford University, Stanford, California 94305, USA}
\end{center}

\vspace{10em}

\newpage
\section{Crystal growth and sample fabrication}
The near optimally doped Bi2212 and Pb-Bi2201 crystals were grown by the floating zone method as described in the previous studies by Howald \textit{et al.~}\cite{howald03}, He \textit{et al.~}\cite{he11}, and references therein. Measurements were performed on BSCCO flakes that were mechanically exfoliated on an oxidized silicon substrate. The flakes' size was chosen smaller than the $\sim$34 \textmu m spot size (beam radius) of the excitation light to ensure uniform illumination, while their thickness was selected to be on the order of 100 nm which is comparable to the optical penetration depth of the 1550 nm wavelength used. For all the devices measured in this study, electrical contacts were patterned using a stencil mask technique in order to reduce degradation of the flakes from exposure to chemicals \cite{sandilands14,zhao19,yu19}. Dried forms of soft polymers such as GE varnish and polymethyl methacrylate were used as shadow masks during metal deposition. \\

\section{Determination of the crystal orientation}
Following the PGE measurements, we determined the crystal orientation of the  Bi2212 flakes using polarized Raman spectroscopy. The Raman spectra were taken in backscattering geometry at room temperature using a 532 nm laser line incident along the $c$ axis of BSCCO. To obtain the orientation dependence of the spectra, the flakes were physically rotated while the polarization axes of the incident and scattered light were fixed parallel to each other.

\begin{figure}[ht]
\includegraphics[width=0.7\columnwidth]{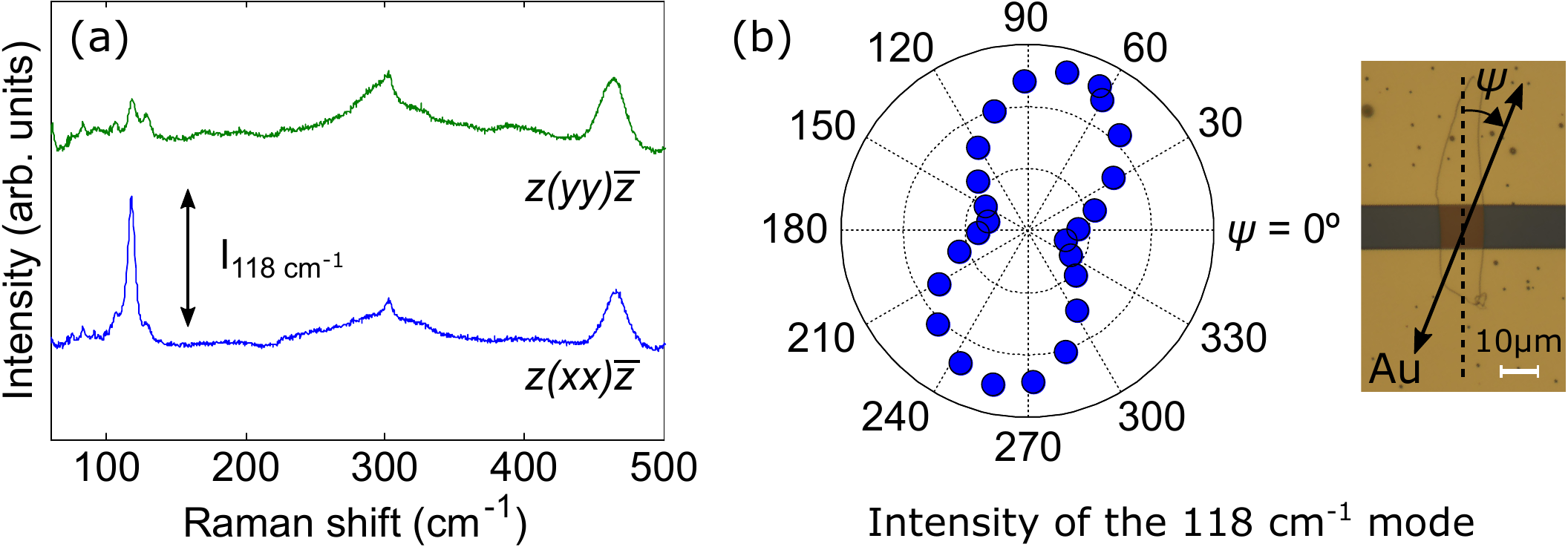}
\caption{Polarized Raman spectra of a 100 nm thick Bi2212 flake. (a) Raman spectra measured in $z(xx)\overline{z}$ and $z(yy)\overline{z}$ scattering configurations. The plots are vertically shifted for clarity. (b) Polar plot of the intensity of the phonon mode at 118 cm\textsuperscript{-1} as a function of $\psi$. As drawn on the optical image, $\psi$ represents the rotation angle between the polarization axis (solid line) and the line that joins the pair of electrical contacts (dashed line).}
\label{raman}
\end{figure}

Figure~\ref{raman}(a) shows the Raman spectra of a Bi2212 flake measured in $z(xx)\overline{z}$ and $z(yy)\overline{z}$ scattering configurations. Here, the $y$ axis denotes the modulation direction of the one-dimensional superstructure, which lies along the diagonals of the CuO\textsubscript{2} square lattice \cite{kirk88}. As reported by the previous Raman studies \cite{cardona88,kirillov88,liu92}, the phonon mode at 118 cm\textsuperscript{-1} displays a strong in-plane anisotropy due to the structural orthorhombicity induced by the superstructure. To see the two-fold anisotropy more clearly, we plot in Fig.~\ref{raman}(b) the mode intensity as a function of the rotation angle of the polarization axis. Based on this intensity pattern, we find that the direction of the current flow in this particular device aligns close to the $y$ axis of Bi2212.

\begin{figure}
\includegraphics[width=0.8\columnwidth]{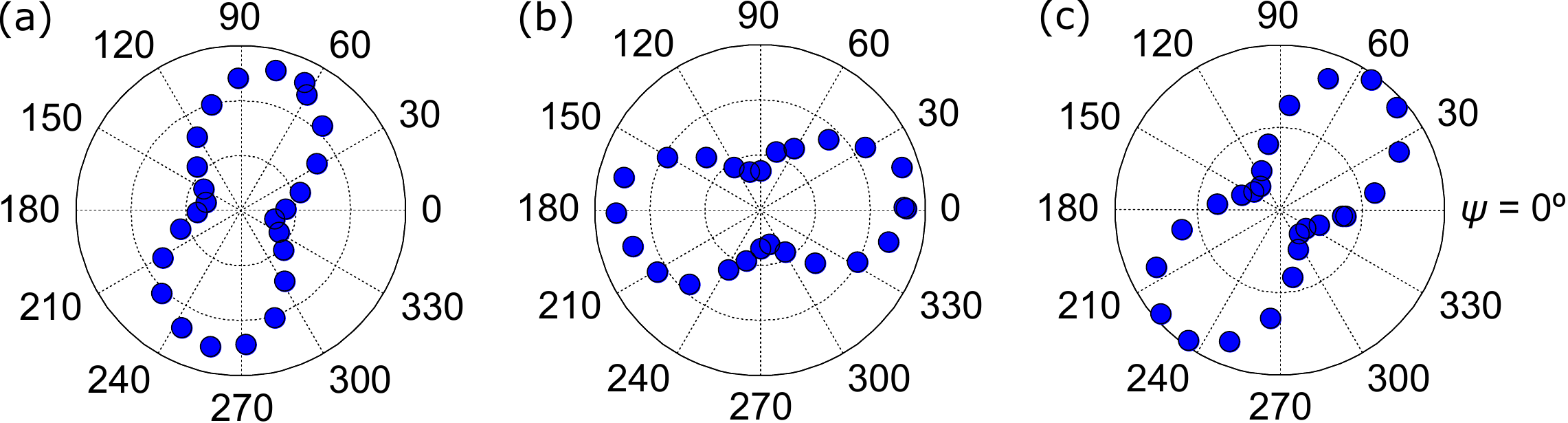}
\caption{Polar plots of the intensity of the phonon mode at 118 cm\textsuperscript{-1} as a function of $\psi$ for the Bi2212 devices presented in Fig.~3(c-h) of the main text.}
\label{orient}
\end{figure}

Figure~\ref{orient} shows the analogous intensity patterns for the Bi2212 devices presented in Fig.~3(c-h) of the main text. The direction of the current flow aligns close to either the (a) $y$ axis, (b) $x$ axis, or (c) Cu-O bond direction.

\section{Raw data and fits to the trivial components}

\begin{figure}[h!]
\includegraphics[width=0.61\columnwidth]{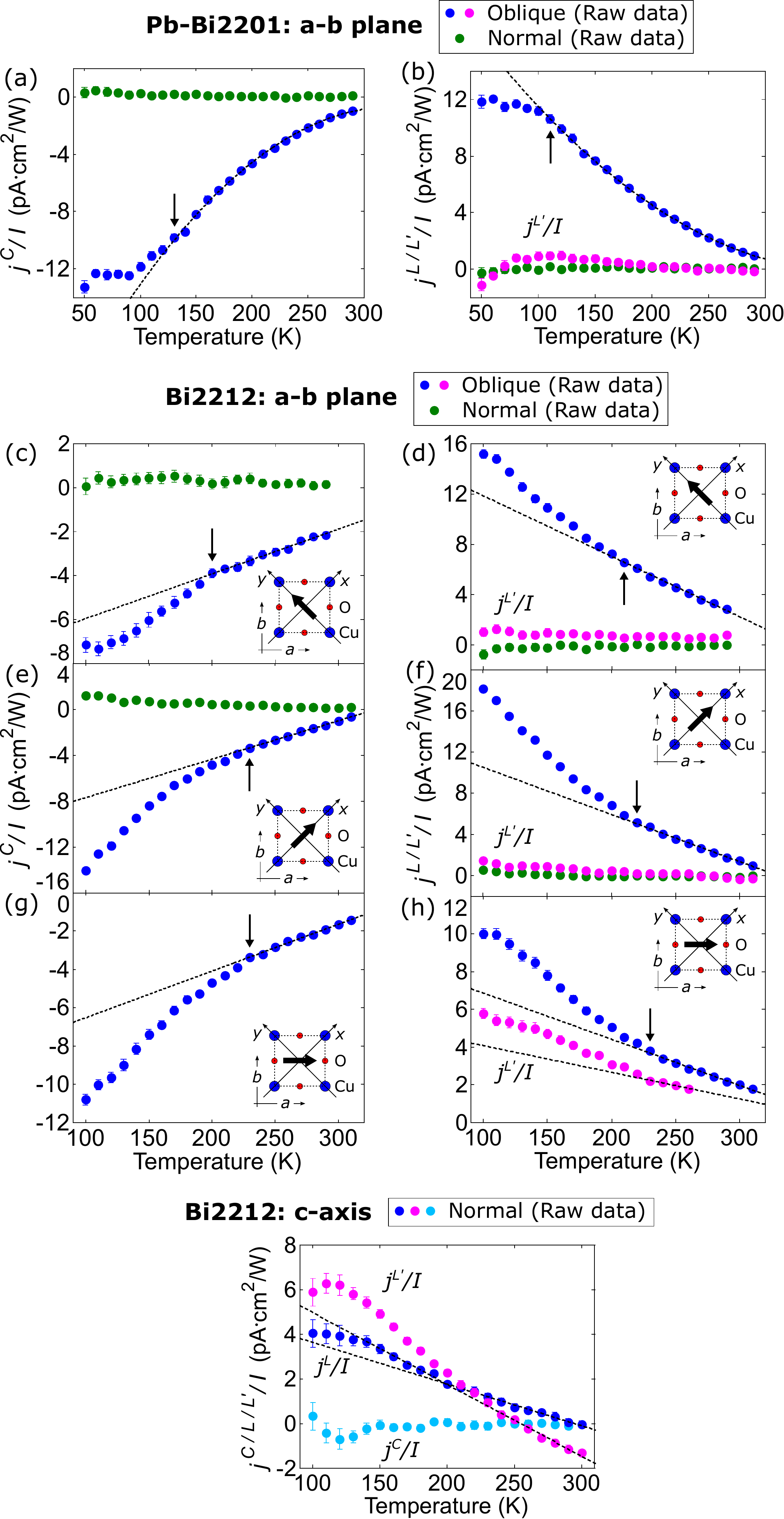}
\caption{Temperature dependence of the fit parameters, normalized by the light intensity $I$, for the (a-b) Pb-Bi2201 and (c-i) Bi2212 devices discussed in the main text. The dashed black lines show fits to the trivial components, and the arrows mark the approximate onset of deviation from the high-temperature trend.}
\label{all}
\end{figure}

As noted in the main text, the photoinduced current is fitted to the phenomenological expression
\be
j = j^{C} \sin 2\varphi + j^{L} \sin 4\varphi + j^{L^\prime} \cos 4\varphi + d.
\label{j}
\ee
Figure \ref{all} shows the temperature dependence of these fit parameters, normalized by the light intensity $I$, for the Pb-Bi2201 and Bi2212 devices discussed in the main text. For $a$-$b$ plane measurements, it is clear that $j^C$ and $j^L$ follow the same temperature dependence but with opposite sign. $j^{L'}$ is typically smaller than $j^C$ and $j^L$, and depends on the orientation of the crystal. When the scattering plane is oriented along the principal plane, either the $x$-$z$ plane (Fig.~\ref{all}(d)) or the $y$-$z$ plane (Fig.~\ref{all}(f)), $j^{L'}$ is much smaller than $j^L$, on the order of $\sim$0.1$j^L$. However, when the plane is rotated away from it (e.g.\ along the diagonal as in Fig.~\ref{all}(h)), $j^{L'}$ becomes finite beyond its typical residual value associated with the photon drag and thermal effects. The constant term $d$ is mainly associated with these accompanying effects. For the $c$ axis measurement, only LPGE seems to be present. The dashed black lines show fits to the trivial components extending down from room temperature.\\

\section{PGE at a simple surface}
For a centrosymmetric material with a simple surface, the surface-induced PGE is characterized by the normal vector $\hat{n}$. From kinetic considerations \cite{Deyo2009}, the CPGE and LPGE currents are then given by
\be
\vec{j}^C\propto \frac{I}{\rho(T)} \hat{n}\times \hat{\Omega} \ \ \ \ \ \  {\rm and} \ \ \ \ \ \ \vec{j}^L\propto \frac{I}{\rho(T)} \hat{e}\times [\hat{e}\times\hat{n}],
\ee
where $\rho$ is the electrical resistivity, $I$ is the light intensity, $\hat{e}$ is the linear component of the light polarization, and $\hat{\Omega}$ is the axial vector associated with the helicity. In the geometry of our experiment (Fig.~1(a) of the main text), where in-plane currents are measured perpendicular to the scattering plane and out-of-plane currents along the surface normal, it is easy to see that obliquely incident radiation generates both the CPGE and LPGE in and out of the plane and normally incident radiation generates only the out-of-plane LPGE.

\section{Symmetries dictated by the experimental observations}

As discussed in the main text, the important features of the nontrivial PGE data below $T^*$ are the following:
\begin{itemize}
\item[\textit{i})] At oblique incidence in the $x$-$z$, $y$-$z$, and diagonal planes, nontrivial CPGE and LPGE are observed in the CuO\textsubscript{2} plane with the following features:
\begin{itemize}
\item Finite currents are measured in the direction perpendicular to the scattering plane;
\item For $j^C$ and $j^L$, the direction of the current flow, either to the left or to the right as seen by looking in the direction of the wave vector $\vec{q}$, is invariant under the in-plane rotation of the crystal;
\item For $j^{L'}$, the current is finite and comparable to $j^L$ only when the scattering plane is rotated away from the high symmetry $x$-$z$ and $y$-$z$ planes;
\end{itemize}
\item[\textit{ii})] At normal incidence, only the nontrivial $c$ axis LPGE is observed.
\end{itemize}
Here, we show in detail how these experimental observations put conditions on the PGE tensor and thus on the symmetry of the pseudogap state. Written as \cite{Deyo2009}
\begin{align}
j_i &= \beta_{ijk} E_j E_k^* \\
&= \chi_{ijk} (E_j E_k^* + E_j^* E_k )/2 + i \gamma_{il} (\vec{E} \times \vec{E}^* )_l, \label{pge_sm} \\
\chi_{ijk}  &= \hbox{Re}[\beta_{ijk}], \\
\gamma_{il} &= \hbox{Im}[\beta_{ijk}] \epsilon_{jkl}/2,
\end{align}
where $\vec{E}$ is the complex amplitude of the electric field, the CPGE is characterized by the second-rank axial ``i''-tensor $\gamma_{il}$, and the LPGE by the third-rank polar ``c''-tensor $\chi_{ijk}$ satisfying $\chi_{ijk} = \chi_{ikj}$ \cite{birss64}.

Let us first find symmetry groups that are consistent with the CPGE data. We then see whether they satisfy the LPGE data as well. \\

\noindent \textbf{CPGE}:

{
\setlength{\leftskip}{10pt}
\setlength{\parindent}{0pt}
Since the symmetry of BSCCO is tetragonal above $T^*$, we limit our analysis to tetragonal, orthorhombic, monoclinic, and triclinic groups.
\begin{itemize}
\item[-] Tetragonal groups: $4\underline{1}$, $4$, $\underline{4}$, $\overline{4}\underline{1}$, $\overline{4}$, $\overline{\underline{4}}$, $4/m\underline{1}$, $4/m$, $\underline{4}/m$, $4/\underline{m}$, $\underline{4}/\underline{m}$, $422\underline{1}$, $422$, $\underline{4}22$, $4\underline{22}$, $4mm\underline{1}$, $4mm$, $\underline{4}m\underline{m}$, $4\underline{mm}$, $\overline{4}2m\underline{1}$, $\overline{4}2m$, $\overline{\underline{4}}2\underline{m}$, $\underline{\overline{4}2}m$, $\overline{4}\underline{2m}$, $4/mmm\underline{1}$, $4/mmm$, $\underline{4}/mm\underline{m}$, $4/m\underline{mm}$, $4/\underline{mmm}$, $4/\underline{m}mm$, and $\underline{4}/\underline{m}m\underline{m}$

Following the notation used by Birss \cite{birss64}, the bars under and over a character denote the time reversal and spatial inversion operations, respectively. The following groups allow the CPGE:
\begin{align*}
4\underline{1},\ 4,\ \underline{4};\ [F_2]:
&\begin{pmatrix}
\gamma_{xx} & \gamma_{xy} & 0 \\
-\gamma_{xy} & \gamma_{xx} & 0 \\
0 & 0 & \gamma_{zz}
\end{pmatrix}, &
\overline{4}\underline{1},\ \overline{4},\ \overline{\underline{4}};\ [G_2]:
&\begin{pmatrix}
\gamma_{xx} & \gamma_{xy} & 0 \\
\gamma_{xy} & -\gamma_{xx} & 0 \\
0 & 0 & 0
\end{pmatrix}, \\
422\underline{1},\ 422,\ \underline{4}22,\ 4\underline{22};\ [H_2]:
&\begin{pmatrix}
\gamma_{xx} & 0 & 0 \\
0 & \gamma_{xx} & 0 \\
0 & 0 & \gamma_{zz}
\end{pmatrix}, &
4mm\underline{1},\ 4mm,\ \underline{4}m\underline{m},\ 4\underline{mm};\ [I_2]:
&\begin{pmatrix}
0 & \gamma_{xy} & 0 \\
-\gamma_{xy} & 0 & 0 \\
0 & 0 & 0
\end{pmatrix}, \\
\overline{4}2m\underline{1},\ \overline{4}2m,\ \overline{\underline{4}}2\underline{m},\ \underline{\overline{4}2}m,\ \overline{4}\underline{2m};\ [J_2]:
&\begin{pmatrix}
\gamma_{xx} & 0 & 0 \\
0 & -\gamma_{xx} & 0 \\
0 & 0 & 0
\end{pmatrix}, & &
\end{align*}
where the tensors are again given in the notation used by Birss \cite{birss64}. For $H_2$ and $J_2$, since $\gamma_{xy} = \gamma_{xz} = \gamma_{yx} = \gamma_{yz} = 0$, the CPGE is not allowed in the direction perpendicular to the scattering plane when light is obliquely incident in either the $x$-$z$ or the $y$-$z$ plane. A similar result is obtained even when we permute the $x$, $y$, and $z$ axes. We therefore rule them out based on observation (\textit{i}). With nonzero off-diagonal components, $G_2$ on the other hand allows for such a current. However, the relation $\gamma_{xy} = \gamma_{yx}$ violates the chirality property. Specifically, the relative direction of current for light incident in the $x$-$z$ plane is opposite of that for light incident in the $y$-$z$ plane. We therefore rule it out also based on observation (\textit{i}). The remaining tensors $F_2$ and $I_2$ are consistent with our data, and this is true only when their principal axes lie along the $c$ axis of BSCCO.

We note that $F_2$ allows for the $c$ axis CPGE at normal incidence of radiation. While we do not observe such a signal in our measurements, we do not consider this symmetry inconsistent with observation (\textit{ii}) since it is possible that our experiment is just not sensitive enough. On a similar note, all the subgroups of allowed high symmetry groups are considered consistent with our data.

Lastly, we apply additional constraints from rotational symmetry breaking by charge order and time reversal symmetry breaking by magnetism. We require the allowed groups to break the four-fold rotational symmetry about the $c$ axis, and exhibit either a paramagnetic or antiferromagnetic behavior or a ferromagnetic behavior with a vanishingly small net magnetic moment. The latter property for ferromagnetism requires that we specify the preferred orientation of magnetic moments. All the symmetry groups that we consider in this study are magnetic in this sense. With these additional constraints, the only possible tetragonal groups are $\underline{4}$ and $\underline{4}m\underline{m}$.

\item[-] Orthorhombic groups: $222\underline{1}$, $222$, $2\underline{22}$, $mm2\underline{1}$, $mm2$, $\underline{mm}2$, $\underline{m}m\underline{2}$, $mmm\underline{1}$, $mmm$, $m\underline{mm}$, $\underline{mmm}$, and $\underline{m}mm$

The following groups allow the CPGE:
\begin{equation*}
222\underline{1},\ 222,\ 2\underline{22};\ [D_2]: 
\begin{pmatrix}
\gamma_{xx} & 0 & 0 \\
0 & \gamma_{yy} & 0 \\
0 & 0 & \gamma_{zz}
\end{pmatrix}, \qquad
mm2\underline{1},\ mm2,\ \underline{mm}2,\ \underline{m}m\underline{2};\ [E_2]:
\begin{pmatrix}
0 & \gamma_{xy} & 0 \\
\gamma_{yx} & 0 & 0 \\
0 & 0 & 0
\end{pmatrix}.
\end{equation*}
We rule out $D_2$ based on observation (\textit{i}). $E_2$ is consistent with our data, provided that the principal axis lies along the $c$ axis of BSCCO. $mm2\underline{1}$ and $mm2$ do not exhibit ferromagnetism in any direction. Although $\underline{mm}2$ and $\underline{m}m\underline{2}$ are ferromagnetic in general, the net moment vanishes when the moments are oriented in and out of the CuO\textsubscript{2} plane, respectively. The allowed orthorhombic groups are then $mm2\underline{1}$, $mm2$, $\underline{mm}2$ (with moments preferentially in the CuO\textsubscript{2} plane), and $\underline{m}m\underline{2}$ (with moments preferentially along the $c$ axis).

\item[-] Monoclinic groups: $2\underline{1}$, $2$, $\underline{2}$, $m\underline{1}$, $m$, $\underline{m}$, $2/m\underline{1}$, $2/m$, $\underline{2}/\underline{m}$, $2/\underline{m}$, and $\underline{2}/m$

The following groups allow the CPGE:
\begin{equation*}
2\underline{1},\ 2,\ \underline{2};\ [B_2]:
\begin{pmatrix}
\gamma_{xx} & \gamma_{xy} & 0 \\
\gamma_{yx} & \gamma_{yy} & 0 \\
0 & 0 & \gamma_{zz}
\end{pmatrix}, \qquad
m\underline{1},\ m,\ \underline{m};\ [C_2]:
\begin{pmatrix}
0 & 0 & \gamma_{xz} \\
0 & 0 & \gamma_{yz} \\
\gamma_{zx} & \gamma_{zy} & 0
\end{pmatrix}.
\end{equation*}
$B_2$ is consistent with our data, provided that the principal axis lies along the $c$ axis of BSCCO. For the ferromagnetic groups $2$ and $\underline{2}$, the net magnetic moment vanishes when the moments are oriented in and out of the plane, respectively. With the principal axis lying along the $c$ axis, $C_2$ shown above violates observation (\textit{i}). But, as can be seen from the same tensor but with permuted axes,
\begin{equation*}
\begin{pmatrix}
0 & \gamma_{xy} & \gamma_{xz} \\
\gamma_{yx} & 0 & 0 \\
\gamma_{zx} & 0 & 0
\end{pmatrix} \quad \hbox{(principal axis along $x$)}, \qquad
\begin{pmatrix}
0 & \gamma_{xy} & 0 \\
\gamma_{yx} & 0 & \gamma_{yz} \\
0 & \gamma_{zy} & 0
\end{pmatrix} \quad \hbox{(principal axis along $y$)},
\end{equation*}
the same symmetry but with the principal axis lying in the CuO\textsubscript{2} plane are consistent with our data.
With the $y$ axis taken as the principal axis, the net magnetic moment vanishes when the moments are oriented in the $x$-$z$ plane for $m$ and along the $y$ axis for $\underline{m}$.

\item[-] Triclinic groups: $1\underline{1}$, $1$, $\overline{1}\underline{1}$, $\overline{1}$, and $\overline{\underline{1}}$

The following groups allow the CPGE:
\begin{equation*}
1\underline{1},\ 1;\ [A_2]:
\begin{pmatrix}
\gamma_{xx} & \gamma_{xy} & \gamma_{xz} \\
\gamma_{yx} & \gamma_{yy} & \gamma_{yz} \\
\gamma_{zx} & \gamma_{zy} & \gamma_{zz}
\end{pmatrix}.
\end{equation*}
There is no principal axis, and $A_2$ with any orientation is consistent with our data.
\end{itemize}

Taken together, the following groups are consistent with our CPGE data:
\begin{itemize}
\item[-] [$I_2$]: $\underline{4}m\underline{m}$
\item[-] [$F_2$]: $\underline{4}$
\item[-] [$E_2$]: $mm2\underline{1}$, $mm2$, $\underline{mm}2$ (with moments preferentially in the CuO\textsubscript{2} plane), $\underline{m}m\underline{2}$ (with moments preferentially along the $c$ axis)
\item[-] [$C_2$ (principal axis along $y$)]: $m\underline{1}$, $m$ (with moments preferentially in the $x$-$z$ plane), $\underline{m}$ (with moments preferentially along the $y$ axis)
\item[-] [$B_2$]: $2\underline{1}$, $2$ (with moments preferentially in the CuO\textsubscript{2} plane), $\underline{2}$ (with moments preferentially along the $c$ axis)
\item[-] [$A_2$ (any orientation)]: $1\underline{1}$, $1$ (with moments preferentially oriented antiparallel to one another)
\end{itemize}
\par}

\noindent \textbf{LPGE}:

{
\setlength{\leftskip}{10pt}
\setlength{\parindent}{0pt}
Now, let us apply to these groups additional constraints from the LPGE data. The LPGE current as defined in Eq.~(\ref{pge_sm}) is in general a sum of terms proportional to $\sin 4\varphi$ and $\cos 4\varphi$ and a constant term. In the general case, represented by the LPGE tensor
\begin{equation*}
\begin{pmatrix}
\chi_{xxx} & \chi_{xyy} & \chi_{xzz} & \chi_{xyz} & \chi_{xxz} & \chi_{xxy} \\
\chi_{yxx} & \chi_{yyy} & \chi_{yzz} & \chi_{yyz} & \chi_{yxz} & \chi_{yxy} \\
\chi_{zxx} & \chi_{zyy} & \chi_{zzz} & \chi_{zyz} & \chi_{zxz} & \chi_{zxy} \\
\end{pmatrix},
\end{equation*}
written in the notation used by Sturman and Fridkin \cite{sturman92}, the coefficients of the $\sin 4\varphi$ and $\cos 4\varphi$ terms are given by
\begin{align}
j_i \left[ \sin 4\varphi \right] &= \frac{I}{4} \left[ \left( \chi_{iyy} - \chi_{ixx} \right) \cos \theta \sin 2\phi + 2 \chi_{ixy} \cos \theta \cos 2\phi + 2 \left( \chi_{ixz} \sin \phi - \chi_{iyz} \cos \phi \right) \sin \theta \right], \label{sin4x} \\
\begin{split}
j_i \left[ \cos 4\varphi \right] &= \frac{I}{4} \left[ \chi_{ixx} \left( \cos^2 \theta \cos^2 \phi - \sin^2 \phi \right) + \chi_{iyy} \left( \cos^2 \theta \sin^2 \phi - \cos^2 \phi \right) + \chi_{izz} \sin^2 \theta \right. \label{cos4x} \\
&\qquad\left. + \chi_{ixy} \left( 1+\cos^2 \theta \right) \sin 2\phi - \left( \chi_{ixz} \cos \phi + \chi_{iyz} \sin \phi \right) \sin 2\theta \right],
\end{split}
\end{align}
for $i = x, y, z$. Here, $\theta$ is the angle between the $c$ axis and the propagation direction of light as illustrated in Fig.~1(a) of the main text, and $\phi$ is the angle between the $x$ axis and the scattering plane. \\

At oblique incidence in the $y$-$z$ plane ($\phi = \pi/2$), for example, the component of the current perpendicular to the scattering plane has the following polarization dependence:
\begin{align}
j_x \left[ \sin 4\varphi \right] &= \frac{I}{2} \left[ \chi_{xxz} \sin \theta - \chi_{xxy} \cos \theta \right], \\
j_x \left[ \cos 4\varphi \right] &= \frac{I}{4} \left[ -\chi_{xxx} + \chi_{xyy} \cos^2 \theta + \chi_{xzz} \sin^2 \theta - \chi_{xyz} \sin 2\theta \right].
\end{align}
As expected, the $\sin 4\varphi$ term requires a field component along the $x$ axis perpendicular to the scattering plane, and the $\cos 4\varphi$ term arises mainly from the field components lying in the plane. \\

Taking into consideration the orientation of the principal axes as determined from the CPGE analysis, the groups consistent with the CPGE data have the following LPGE tensors:
\begin{align*}
\underline{4}m\underline{m};\ [(J_3)]&:
\begin{pmatrix}
0 & 0 & 0 & 0 & \chi_{xxz} & 0 \\
0 & 0 & 0 & -\chi_{xxz} & 0 & 0 \\
\chi_{zxx} & -\chi_{zxx} & 0 & 0 & 0 & 0
\end{pmatrix}, \\
\underline{4};\ [G_3]&:
\begin{pmatrix}
0 & 0 & 0 & \chi_{xyz} & \chi_{xxz} & 0 \\
0 & 0 & 0 & -\chi_{xxz} & \chi_{xyz} & 0 \\
\chi_{zxx} & -\chi_{zxx} & 0 & 0 & 0 & \chi_{zxy}
\end{pmatrix}, \\
mm2\underline{1},\ mm2;\ [E_3]&:
\begin{pmatrix}
0 & 0 & 0 & 0 & \chi_{xxz} & 0 \\
0 & 0 & 0 & \chi_{yyz} & 0 & 0 \\
\chi_{zxx} & \chi_{zyy} & \chi_{zzz} & 0 & 0 & 0
\end{pmatrix}, \\
\underline{mm}2;\ [D_3]&:
\begin{pmatrix}
0 & 0 & 0 & \chi_{xyz} & 0 & 0 \\
0 & 0 & 0 & 0 & \chi_{yxz} & 0 \\
0 & 0 & 0 & 0 & 0 & \chi_{zxy}
\end{pmatrix}, \\
\underline{m}m\underline{2};\ [(E_3)]&:
\begin{pmatrix}
0 & 0 & 0 & 0 & 0 & \chi_{xxy} \\
\chi_{yxx} & \chi_{yyy} & \chi_{yzz} & 0 & 0 & 0 \\
0 & 0 & 0 & \chi_{zyz} & 0 & 0
\end{pmatrix}, \\
m\underline{1},\ m;\ [C_3 \hbox{ with principal axis along } y]&:
\begin{pmatrix}
\chi_{xxx} & \chi_{xyy} & \chi_{xzz} & 0 & \chi_{xxz} & 0 \\
0 & 0 & 0 & \chi_{yyz} & 0 & \chi_{yxy} \\
\chi_{zxx} & \chi_{zyy} & \chi_{zzz} & 0 & \chi_{zxz} & 0
\end{pmatrix}, \\
\underline{m};\ [B_3 \hbox{ with principal axis along } y]&:
\begin{pmatrix}
0 & 0 & 0 & \chi_{xyz} & 0 & \chi_{xxy} \\
\chi_{yxx} & \chi_{yyy} & \chi_{yzz} & 0 & \chi_{yxz} & 0 \\
0 & 0 & 0 & \chi_{zyz} & 0 & \chi_{zxy}
\end{pmatrix}, \\
2\underline{1},\ 2;\ [B_3]&:
\begin{pmatrix}
0 & 0 & 0 & \chi_{xyz} & \chi_{xxz} & 0 \\
0 & 0 & 0 & \chi_{yyz} & \chi_{yxz} & 0 \\
\chi_{zxx} & \chi_{zyy} & \chi_{zzz} & 0 & 0 & \chi_{zxy}
\end{pmatrix}, \\
\underline{2};\ [C_3]&:
\begin{pmatrix}
\chi_{xxx} & \chi_{xyy} & \chi_{xzz} & 0 & 0 & \chi_{xxy} \\
\chi_{yxx} & \chi_{yyy} & \chi_{yzz} & 0 & 0 & \chi_{yxy} \\
0 & 0 & 0 & \chi_{zyz} & \chi_{zxz} & 0
\end{pmatrix}, \\
1\underline{1},\ 1;\ [A_3]&:
\begin{pmatrix}
\chi_{xxx} & \chi_{xyy} & \chi_{xzz} & \chi_{xyz} & \chi_{xxz} & \chi_{xxy} \\
\chi_{yxx} & \chi_{yyy} & \chi_{yzz} & \chi_{yyz} & \chi_{yxz} & \chi_{yxy} \\
\chi_{zxx} & \chi_{zyy} & \chi_{zzz} & \chi_{zyz} & \chi_{zxz} & \chi_{zxy} \\
\end{pmatrix}.
\end{align*}
First, we apply to these tensors constraints from the $\sin 4\varphi$ data. At oblique incidence, we require the current to appear perpendicular to all the high symmetry planes and satisfy the chirality property (observation (\textit{i})). At normal incidence, we require the current to appear along the $c$ axis (observation (\textit{ii})). Using Eq.~(\ref{sin4x}), we find that $E_3$, $C_3$ (with the principal axis along $y$), $B_3$, and $A_3$ are consistent with these observations. Besides these groups, $C_3$ with nonzero $\chi_{xxy}$ and $\chi_{yxy}$ components, in principle, also allows the $\sin 4\varphi$ term for light incident in the $x$-$z$ and $y$-$z$ planes. However, our complementary observation of the negligible in-plane LPGE at normal incidence indicates that these tensor components are small and cannot account for the finite $\sin 4\varphi$ term observed at oblique incidence. We therefore consider this tensor inconsistent with our LPGE data. The symmetry groups consistent with both the CPGE and LPGE $\sin 4\varphi$ data are then $mm2\underline{1}$, $mm2$, $m\underline{1}$, $m$ (with the principal axis along $y$, and magnetic moments preferentially in the $x$-$z$ plane), $2\underline{1}$, $2$ (with moments preferentially in the CuO\textsubscript{2} plane), $1\underline{1}$, and $1$ (with moments preferentially oriented antiparallel to one another). \\

So far, our analysis was based on finite signals that we observe experimentally. Upon comparison with predictions, the symmetry groups that forbid such signals in disagreement with our measurements were considered inconsistent and therefore ruled out. An alternative approach is based on small signals that we cannot resolve clearly. Since it is possible that the signals are finite but too weak to be detected by our experiment, requiring the symmetry groups to forbid such signals is a strong statement. However, such an analysis provides useful insights, and we now apply it to our LPGE $\cos 4\varphi$ data.  \\

As discussed in the main text and also in the section above on various components of the photocurrent, the $\cos 4\varphi$ term of the LPGE is small on the order of $\sim$0.1$j^L$ when the scattering plane is oriented along the $x$-$z$ and $y$-$z$ planes, and becomes finite and comparable to $j^L$ when the scattering plane is rotated away from these high symmetry planes. Interestingly, as can be seen by applying Eq.~(\ref{cos4x}), $mm2\underline{1}$ and $mm2$ predict such an anisotropic effect by forbidding the $\cos 4\varphi$ current perpendicular to the $x$-$z$ and $y$-$z$ planes. The other groups $m\underline{1}$, $m$, $2\underline{1}$, $2$, $1\underline{1}$, and $1$ allow such a current, although the predicted current may be small. With this additional constraint from the LPGE $\cos 4\varphi$ data, the only symmetry groups consistent with both the CPGE and LPGE data are $mm2\underline{1}$ and $mm2$. \\

Table \ref{summary} summarizes our analysis. Figure \ref{mm2} illustrates spin structures consistent with $mm2\underline{1}$ and $mm2$.

\begin{table}
\begin{center}
\scriptsize
\def\arraystretch{1.8}
\begin{tabular}{| c | c | c | c | c | c | c | c | c | c | c | c | c | c | c |}
\hline
\multicolumn{4}{|c|}{ \multirow{2}{*}{Group properties}} & \multicolumn{3}{c|}{$j^C$} & \multirow{2}{4.6em}{\centering Rotation symmetry} & \multicolumn{3}{c|}{$j^L$} & \multicolumn{4}{c|}{$j^{L'}$} \\
\cline{5-7} \cline{9-15}
\multicolumn{4}{|c|}{} & \multicolumn{3}{c|}{Oblique} & & \multicolumn{2}{c|}{Oblique} & Normal & \multicolumn{4}{c|}{Oblique} \\
\hline
Group & \multicolumn{1}{m{4.4em}|}{\centering Principal axis} & CPGE & LPGE & \multicolumn{1}{m{4.6em}|}{\centering Allows $j_\perp$ for all high symmetry planes} & Is chiral & \multicolumn{1}{m{5.3em}|}{\centering Crystal orientation of principal axis} & Lacks $4_z$ & \multicolumn{1}{m{4.6em}|}{\centering Allows $j_\perp$ for all high symmetry planes} & Is chiral & \multicolumn{1}{m{4.6em}|}{\centering Allows $j_c$ for an arbitrary symmetry plane} & \multicolumn{1}{m{2.9em}|}{\centering Allows $j_\perp$ for $x$-$z$ plane} & \multicolumn{1}{m{2.9em}|}{\centering Allows $j_\perp$ for $y$-$z$ plane} & \multicolumn{1}{m{2.9em}|}{\centering Allows $j_\perp$ for [$xy$]-$z$ plane} & \multicolumn{1}{m{3.1em}|}{\centering Allows $j_\perp$ for [-$xy$]-$z$ plane} \\
\hline \hline
$4\underline{1}$ & $4_z$ & $F_2$ & $F_3$ & \yes & \yes & $c$ & \no & \yes & \yes & \no & \yes & \yes & \yes & \yes \\
\hline
$4$ & $4_z$ & $F_2$ & $F_3$ & \yes & \yes & $c$ & \no & \yes & \yes & \no & \yes & \yes & \yes & \yes \\
\hline
$\underline{4}$ & $\underline{4}_z$ & $F_2$ & $G_3$ & \yes & \yes & $c$ & \yes & \yes & \no & \yes & \yes & \yes & \yes & \yes \\
\hline
$\overline{4}\underline{1}$ & $\overline{4}_z$ & $G_2$ & $G_3$ & \yes & \no & $c$ & \yes & \yes & \no & \yes & \yes & \yes & \yes & \yes \\
\hline
$\overline{4}$ & $\overline{4}_z$ & $G_2$ & $G_3$ & \yes & \no & $c$ & \yes & \yes & \no & \yes & \yes & \yes & \yes & \yes \\
\hline
$\underline{\overline{4}}$ & $\underline{\overline{4}}_z$ & $G_2$ & $F_3$ & \yes & \no & $c$ & \yes & \yes & \yes & \no & \yes & \yes & \yes & \yes \\
\hline
$422\underline{1}$ & $4_z$ & $H_2$ & $H_3$ & \no & \na & \na & \no \\
\cline{1-8}
$422$ & $4_z$ & $H_2$ & $H_3$ & \no & \na & \na & \no \\
\cline{1-8}
$\underline{4}22$ & $\underline{4}_z$ & $H_2$ & $J_3$ & \no & \na & \na & \yes \\
\cline{1-8}
$4\underline{22}$ & $4_z$ & $H_2$ & $I_3$ & \no & \na & \na & \no \\
\hline
$4mm\underline{1}$ & $4_z$ & $I_2$ & $I_3$ & \yes & \yes & $c$ & \no & \yes & \yes & \no & \no & \no & \no & \no \\
\hline
$4mm$ & $4_z$ & $I_2$ & $I_3$ & \yes & \yes & $c$ & \no & \yes & \yes & \no & \no & \no & \no & \no \\
\hline
$\underline{4}m\underline{m}$ & $\underline{4}_z$ & $I_2$ & ($J_3$) & \yes & \yes & $c$ & \yes & \no & \na & \yes & \no & \no & \yes & \yes \\
\hline
$4\underline{mm}$ & $4_z$ & $I_2$ & $H_3$ & \yes & \yes & $c$ & \no & \no & \na & \no & \yes & \yes & \yes & \yes \\
\hline
$\overline{4}2m\underline{1}$ & $\overline{4}_z$ & $J_2$ & $J_3$ & \no & \na & \na & \yes \\
\cline{1-8}
$\overline{4}2m$ & $\overline{4}_z$ & $J_2$ & $J_3$ & \no & \na & \na & \yes \\
\cline{1-8}
$\underline{\overline{4}}2\underline{m}$ & $\underline{\overline{4}}_z$ & $J_2$ & $H_3$ & \no & \na & \na & \yes \\
\cline{1-8}
$\underline{\overline{4}2}m$ & $\underline{\overline{4}}_z$ & $J_2$ & $I_3$ & \no & \na & \na & \yes \\
\cline{1-8}
$\overline{4}\underline{2m}$ & $\overline{4}_z$ & $J_2$ & ($J_3$) & \no & \na & \na & \yes \\
\cline{1-8}
$222\underline{1}$ & $2_x$, $2_y$, $2_z$ & $D_2$ & $D_3$ & \no & \na & \na & \yes \\
\cline{1-8}
$222$ & $2_x$, $2_y$, $2_z$ & $D_2$ & $D_3$ & \no & \na & \na & \yes \\
\cline{1-8}
$2\underline{22}$ & $2_z$ & $D_2$ & $E_3$ & \no & \na & \na & \yes \\
\hline
$mm2\underline{1}$ & $2_z$ & $E_2$ & $E_3$ & \yes & \yes & $c$ & \yes & \yes & \yes & \yes & \no & \no & \yes & \yes \\
\hline
$mm2$ & $2_z$ & $E_2$ & $E_3$ & \yes & \yes & $c$ & \yes & \yes & \yes & \yes & \no & \no & \yes & \yes \\
\hline
$\underline{mm}2$ & $2_z$ & $E_2$ & $D_3$ & \yes & \yes & $c$ & \yes & \no & \na & \yes & \yes & \yes & \yes & \yes \\
\hline
$\underline{m}m\underline{2}$ & $\underline{2}_z$ & $E_2$ & ($E_3$) & \yes & \yes & $c$ & \yes & \no & \na & \no & \yes & \no & \yes & \yes \\
\hline
$2\underline{1}$ & $2_z$ & $B_2$ & $B_3$ & \yes & \yes & $c$ & \yes & \yes & \yes & \yes & \yes & \yes & \yes & \yes \\
\hline
$2$ & $2_z$ & $B_2$ & $B_3$ & \yes & \yes & $c$ & \yes & \yes & \yes & \yes & \yes & \yes & \yes & \yes \\
\hline
$\underline{2}$ & $\underline{2}_z$ & $B_2$ & $C_3$ & \yes & \yes & $c$ & \yes & \maybe & \na & \no & \yes & \yes & \yes & \yes \\
\hline
$m\underline{1}$ & $\overline{2}_z$ & $C_2$ & $C_3$ & \yes & \yes & $x$ or $y$ & \yes & \yes & \yes & \yes & \no & \yes & \yes & \yes \\
\hline
$m$ & $\overline{2}_z$ & $C_2$ & $C_3$ & \yes & \yes & $x$ or $y$ & \yes & \yes & \yes & \yes & \no & \yes & \yes & \yes \\
\hline
$\underline{m}$ & $\underline{\overline{2}}_z$ & $C_2$ & $B_3$ & \yes & \yes & $x$ or $y$ & \yes & \no & \na & \yes & \yes & \yes & \yes & \yes \\
\hline
$1\underline{1}$ & None & $A_2$ & $A_3$ & \yes & \yes & Any & \yes & \yes & \yes & \yes & \yes & \yes & \yes & \yes \\
\hline
$1$ & None & $A_2$ & $A_3$ & \yes & \yes & Any & \yes & \yes & \yes & \yes & \yes & \yes & \yes & \yes \\
\hline
\end{tabular}
\end{center}
\caption{ Summary of the analysis. Out of all the tetragonal, orthorhombic, monoclinic, and triclinic groups, only the ones that allow both the CPGE and LPGE are shown. All the symmetry groups are considered to be magnetic: (paramagnetic, antiferromagnetic, or ferromagnetic.) The $j^L$ and $j^{L'}$ terms were analyzed only for those groups with known orientation of the principal axes as determined from the CPGE data. The $\checkmark$ mark denotes that the condition is satisfied; the $\times$ mark denotes that the condition is not satisfied; the $-$ mark denotes that the condition cannot be tested; and the $\bigtriangleup$ marks denotes that the condition is satisfied in principle but the effect is likely small based on complementary measurements.}
\label{summary}
\end{table}

\begin{figure}[ht]
\centering
\subfloat[$mm2\underline{1}$]{\includegraphics[width=0.25\columnwidth]{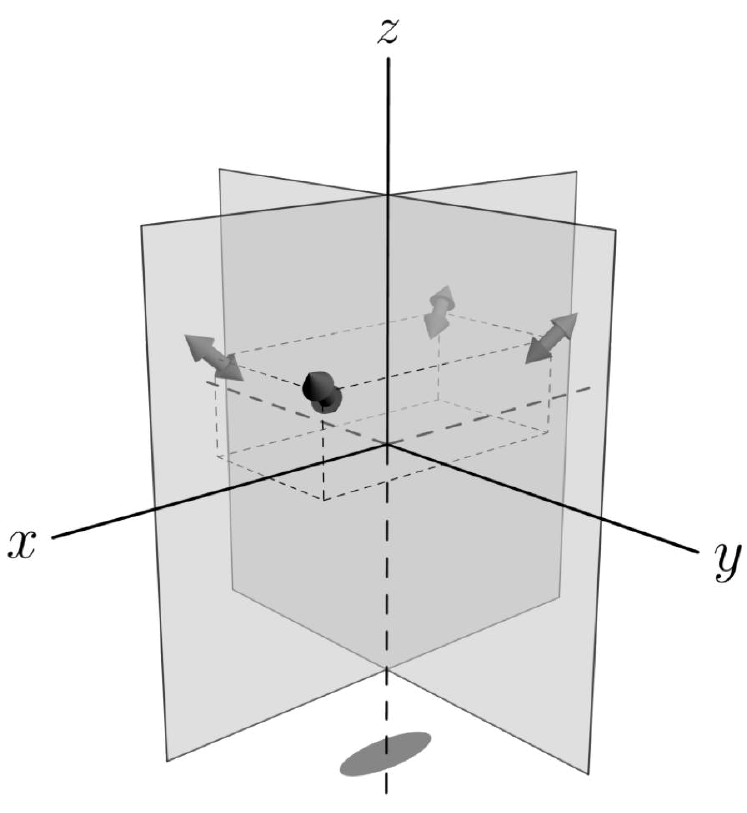}}
%\hspace{3em}
\subfloat[$mm2$]{\includegraphics[width=0.25\columnwidth]{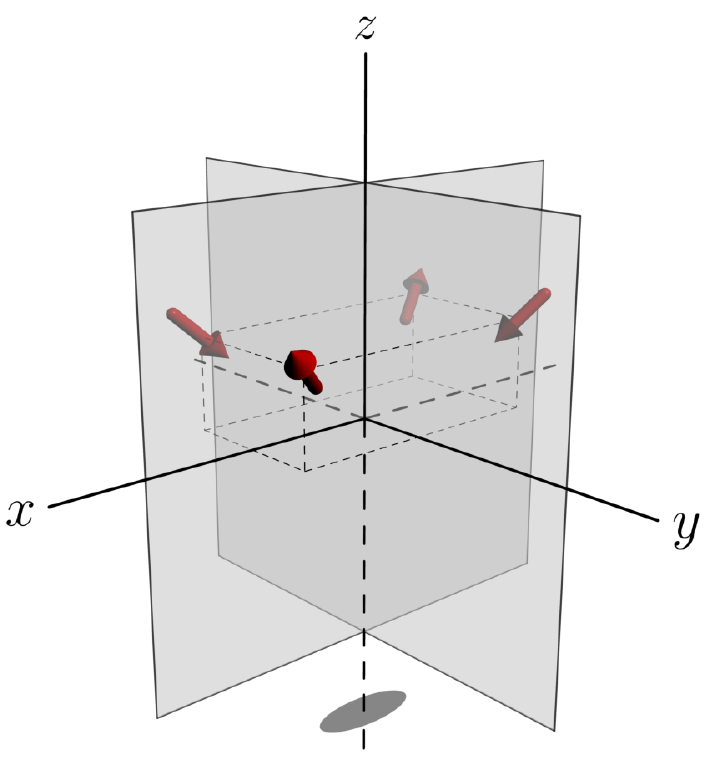}}
%\hspace{3em}
\subfloat[$mm2-loop~current$]{\includegraphics[width=0.25\columnwidth]{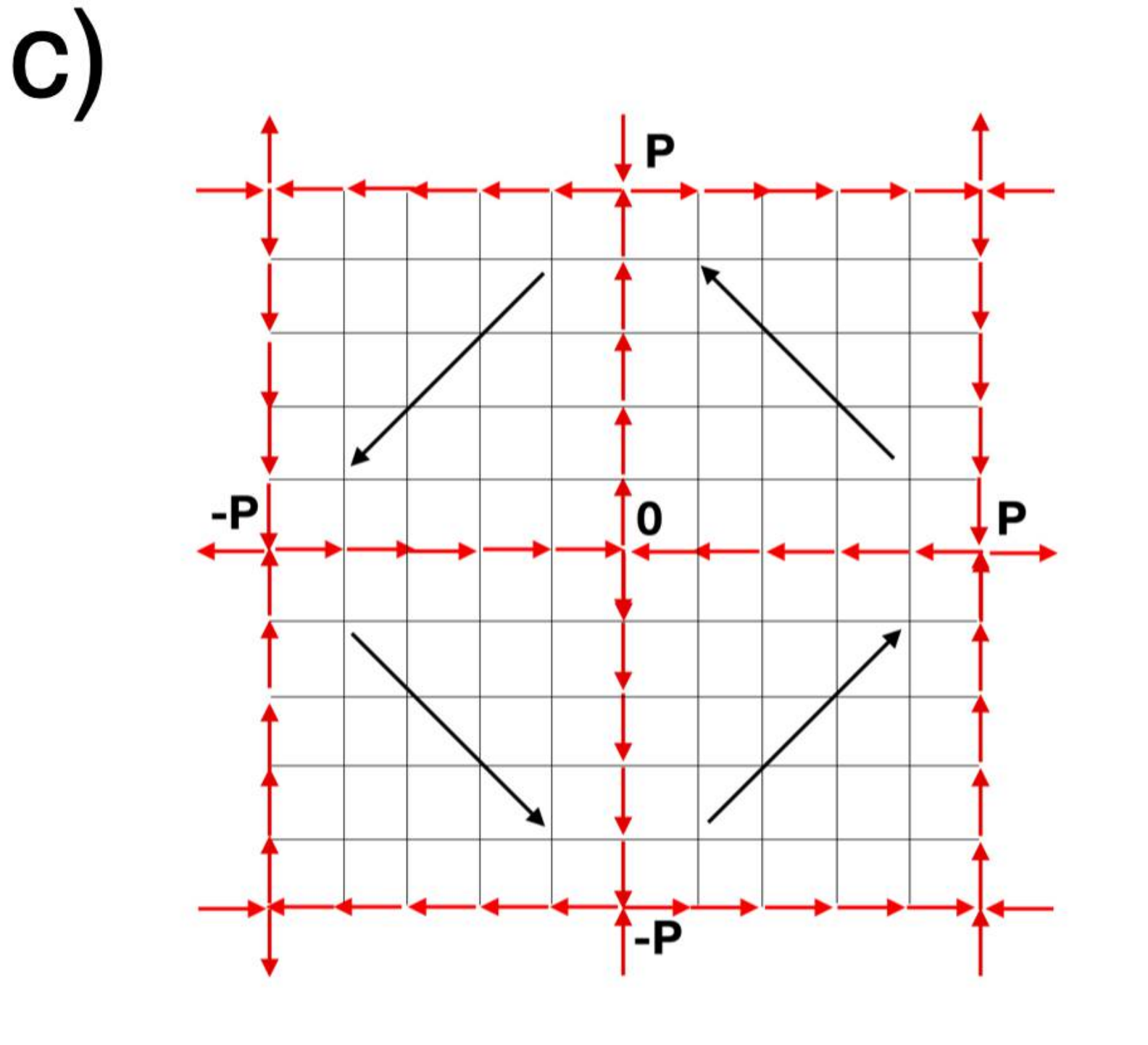}}
\caption{Illustration of spin structures consistent with the (a) $mm2\underline{1}$, (b) $mm2$, (c) The four domains of loop-current order, shown by the direction of the anapole order parameter, arranged as shown in a supr-cell with 2P X 2P original cells, as suggested in Ref.~\cite{varma2019}. P=5 cells of each domain is shown with boundary currents forming a topological flux pattern.}
	\label{mm2}
\end{figure}

\begin{figure}[ht]
\centering
\subfloat[$m\underline{1}$ (principal axis along $y$)]{\includegraphics[width=0.25\columnwidth]{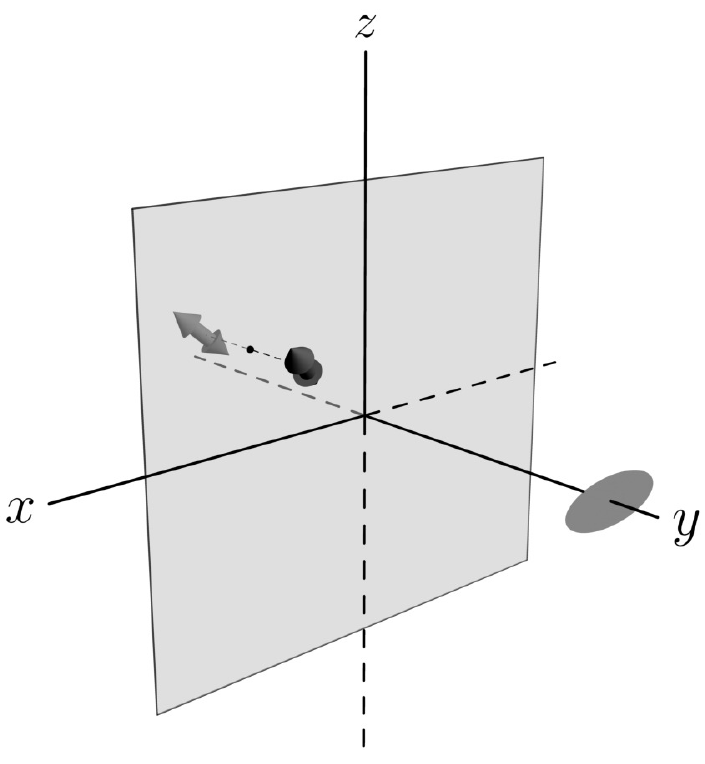}}
\hspace{7em}
\subfloat[$m$ (principal axis along $y$)]{\includegraphics[width=0.25\columnwidth]{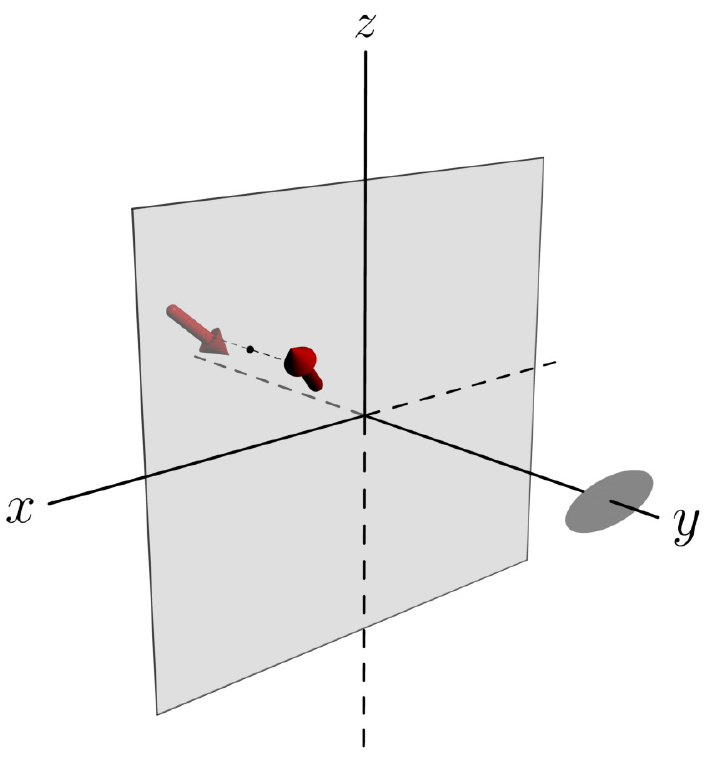}}
\\
\subfloat[$2\underline{1}$]{\includegraphics[width=0.25\columnwidth]{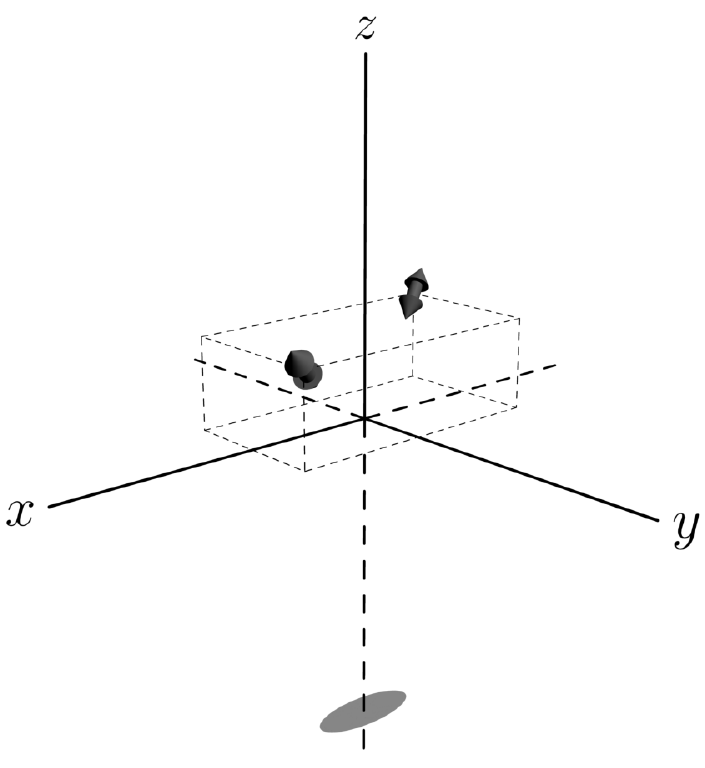}}
\hspace{7em}
\subfloat[$2$]{\includegraphics[width=0.25\columnwidth]{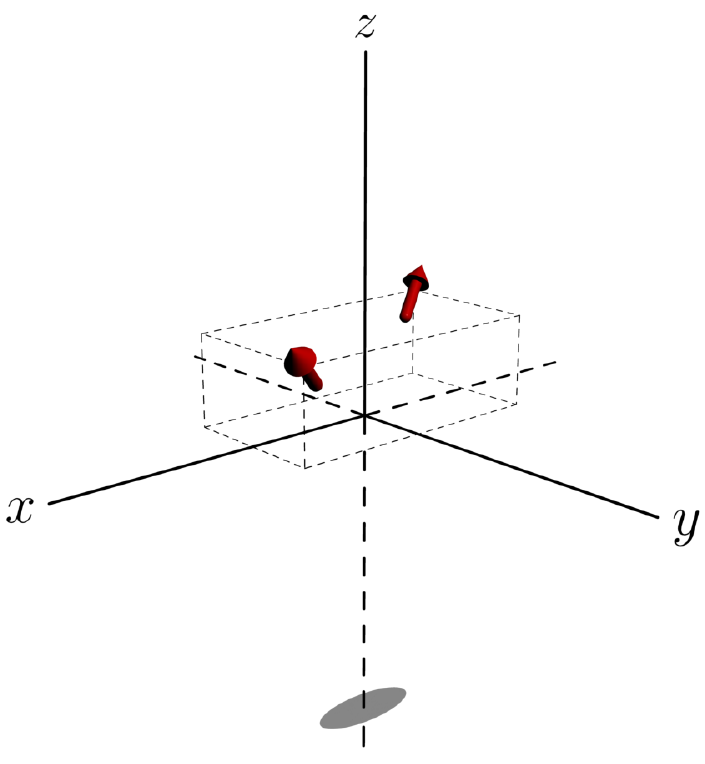}}
\\
\subfloat[$\underline{2}$]{\includegraphics[width=0.25\columnwidth]{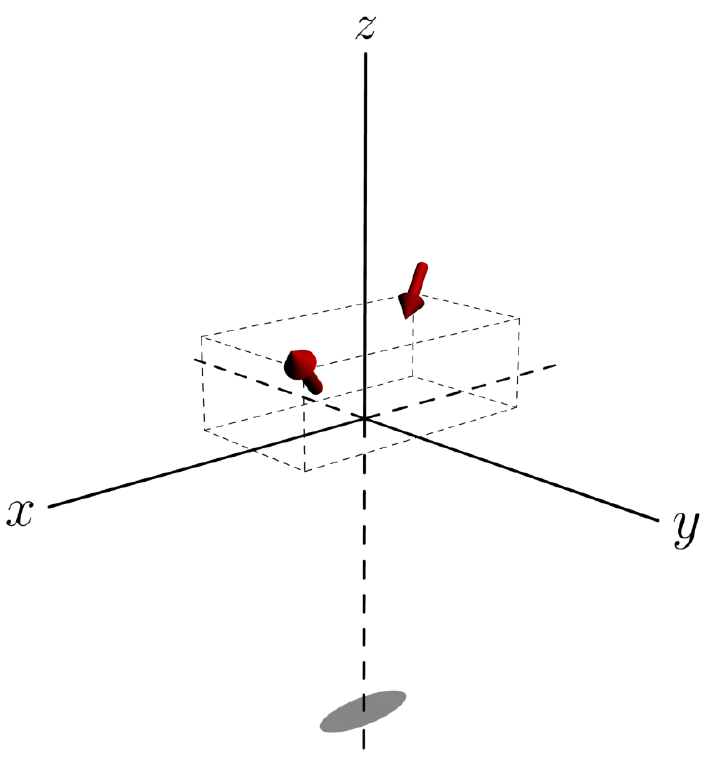}}
\hspace{7em}
\subfloat[$\underline{4}$]{\includegraphics[width=0.25\columnwidth]{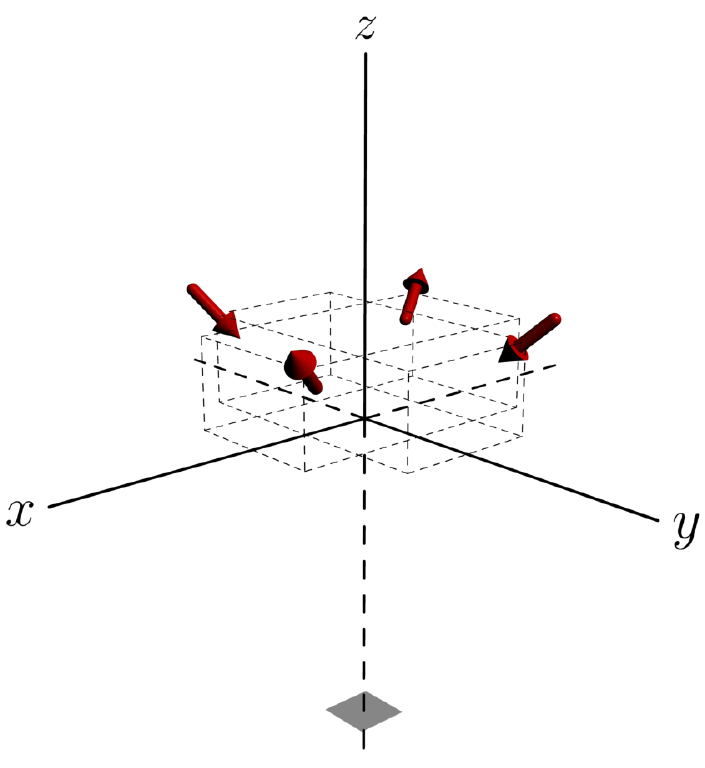}}
\caption{(a) $m\underline{1}$ (principal axis along $y$), (b) $m$ (principal axis along $y$), (c) $2\underline{1}$, (d) $2$, (e) $\underline{2}$, and (f) $\underline{4}$ symmetry groups. The ellipse and square drawn along the axes denote the two-fold and four-fold rotational symmetries, respectively. The gray planes denote the mirror planes.}
\label{mx}
\end{figure}

\section{A possible model discussed in the text}

Recently, in order to understand the phenomena of Fermi-arcs and small Fermi-surface magneto-oscillations in the pseudogap phase absent in ${m\underline{m}m}$,
 a modification of the loop-current order $m\underbar{m}m$ symmetry \cite{simon02}, was proposed \cite{varma2019}.  The modification consists simply of the topological arrangements obtained by gluing the four domains of $m\underline{m}m$ subject to the absolute requirement that currents at the domain boundaries are conserved. This is sketched in Fig.~\ref{mm2}(c); this arrangement has the symmetry representations of $mm2$ listed above, considering the new much larger unit-cell, as one must. This symmetry with domains of size of about 4 or more lattice constants would be consistent with neutron scattering and its monoclinic analog with the SHG. $mm2$ with ${\underline{m}}mm$ as the underlying structure in the unit-cells as described in \cite{simon02}, has the most intense magnetic Bragg spots of the latter in polarized neutron scattering and superstructure Bragg spots with intensity only of $\mathcal{O}(a_0/d)^2$. 
SHG and several other experiments are proportional to quadratic order in the parameters while our experiment is proportional to the order parameter.  It is therefore much very sensitive to the domains which if random would average the PGE current to $\mathcal{O}(d/L)$, where 
$d$ is the typical size of the domains and $L$ the size of the illuminated region $\mathcal{O}(0.1 mm)$, i.e. about  $2 \times 10^4 a_0$, where $a_0$ is the lattice constant. Neutron scattering in Bi2212 \cite{dealmeidadidry12} observes the presence of domains and estimates a correlation of only about 3 lattice constants. Random arrangement of domains would then make the effect unobservable. An ordered arrangement of domains (necessary for orbital current orders) reduces the effect only by $O((a/d)^2)$ compared to the case that each original unit-has
$mm2$ symmetry. Any disorder or charge density waves coupling to the domain orientations preserves the topology of the domains as emphasized in Ref. (\cite{varma2019}) but they also reduce the magnitude of the effects.  Other aspects preserved in such a 
$mm2$ symmetry are that its quantum critical fluctuations have the scale invariance which gives the marginal fermi-liquid properties of the strange metal phase \cite{varma-rmp2020}.

\end{document}